\documentclass[10pt,journal, draftcls]{IEEEtran}

\onecolumn

\usepackage{mathptmx}
\usepackage{times} 
\usepackage{amsmath} 
\usepackage{amsbsy} 
\usepackage{amssymb} 
\usepackage{mathrsfs}
\usepackage{comment}
\usepackage[export]{adjustbox}
\usepackage{tikz}
\usetikzlibrary{external,positioning,decorations.pathreplacing,shapes,arrows,patterns}
\usepackage[outline]{contour}

\usepackage{pgfplots}
\usepackage{graphicx}
\usepackage{pstool}
\usepackage[latin1]{inputenc}
\usepackage{xifthen}
\usepackage{epic}
\usepackage{subcaption}
\usepackage{caption}
\usepackage[outdir=./]{epstopdf}

\newtheorem{thm}{\bf{Theorem}}
\newtheorem{cor}[thm]{\bf {Corollary}}
\newtheorem{lem}[thm]{\bf {Lemma}}
\newtheorem{prop}[thm]{\bf {Proposition}}

\newcommand{\mmse}{\mathsf{mmse}}

\newcommand{\enc}{\mathsf{enc}}
\newcommand{\dec}{\mathsf{dec}}
\newcommand{\smp}{\mathsf{smp}}
\newcommand{\qnt}{\mathsf{qnt}}

\newcommand{\CE}{\mathrm{CE} }
\newcommand{\Ltwo}{\mathrm{L}_2 }
\newcommand{\OPTA}{\mathrm{OPTA} }

\newcommand{\intzerotoone}{\int_0^1}
\newcommand{\Um}{\mathbf U}

\tikzstyle{int}=[draw, fill=blue!10, minimum height = 1cm, minimum width=1.5cm,thick ]
\tikzstyle{sint}=[draw, fill=blue!10, minimum height = 0.5cm, minimum width=0.8cm,thick ]
\tikzstyle{sum}=[circle, fill=blue!10, draw=black,line width=1pt,minimum size = 0.5cm, thick ]
\tikzstyle{ssum}=[circle, fill=blue!10,draw=black,line width=1pt,minimum size = 0.1cm]
\tikzstyle{int1}=[draw, fill=blue!10, minimum height = 0.5cm, minimum width=1cm,thick ]
\tikzstyle{enc}=[draw, fill=blue!10, minimum height = 2.7cm, minimum width=1cm,thick ]
\tikzstyle{int}=[draw, fill=blue!10, minimum height = 1cm, minimum width=1.5cm,thick ]

\title{\LARGE \bf
The Distortion-Rate Function of Sampled Wiener Processes}

\author{ 
\IEEEauthorblockN{
Alon Kipnis,  Andrea J. Goldsmith and Yonina C. Eldar}

\thanks{ A. Kipnis is with the Department of Statistics, Stanford University, Stanford, CA 94305 USA. 
A. J. Goldsmith are with the Department of Electrical Engineering, Stanford University, Stanford, CA 94305 USA. 
Y. C. Eldar is with the Department of Electrical Engineering, Technion - Israel Institute of Technology, Haifa 32000, Israel.}

\thanks{This paper was presented in part at the International Symposium on Information Theory, Barcelona, Spain, 2016 \cite{WienerISIT}. }

}

\begin{document}
\graphicspath{{../Figures/}}
\maketitle

\begin{abstract}
We consider the recovery of a continuous-time Wiener process from a quantized or lossy compressed version of its uniform samples under limited bitrate and sampling rate. We derive a closed form expression for the optimal tradeoff among sampling rate, bitrate, and quadratic distortion in this setting. This expression is given in terms of a reverse waterfilling formula over the asymptotic spectral distribution of a sequence of finite-rank operators associated with the optimal estimator of the Wiener process from its samples. We show that the ratio between this expression and the standard distortion rate function of the Wiener process, describing the optimal tradeoff between bitrate and distortion without a sampling constraint, is only a function of the number of bits per sample. For example using one bit per sample on average, the expected distortion is approximately $1.2$ times the standard distortion rate function, indicating a performance loss of about $20\%$ due to sampling. We next consider the distortion when the continuous-time process is estimated from the output of an encoder that is optimal with respect to the discrete-time samples. 
We show that while the latter is strictly greater than the distortion under optimal encoding, the ratio between the two does not exceed $1.027$. We therefore conclude that nearly optimal performance is attained even if the encoder is unaware of the sampling rate and encodes the samples without taking into account the continuous-time underlying process. \end{abstract}


\section{Introduction}
\
The Wiener process is a Gaussian stochastic process with stationary independent increments and continuous sample paths, with extensive applications in theoretical and applied science. In particular, 
the Wiener process models motion of diffusion particles, it is the driving process of risky financial assets in financial mathematics \cite{karatzas1998methods}, it arises as the limiting law of sequential hypotesis testing procedures \cite{siegmund2013sequential}, it provides the basis for continuous-time martingale theory \cite{karatzas1991brownian}, and it is used to model phase noise in some communication channels \cite{foschini1988characterizing}. In this work we are concerned with the problem of encoding the path of a Wiener process using a limited number of bits per unit time (bitrate). This is a source coding (lossy compression) problem that arises when a random signal whose probability law follows that of the Wiener process is stored or processed in digital memory, or transmitted over a link of limited capacity. \\

The \emph{distortion-rate function} (DRF) of the Wiener process describes the optimal trade-off between bitrate and distortion in its encoding and reconstruction. This DRF was derived by Berger in \cite{berger1970information}, and is based on an encoding of the coefficients of the Karhunen-L{\`o}eve (KL) expansion of the Wiener process. 
These coefficients are obtained by integrating the Wiener path with respect to the KL basis elements. In parctice, however, implementing such integration using purely analog components is extremely challenging since the Wiener process has equal energy in all its frequency components, whereas electronic devices tend to attenuate high frequencies. Consequently, algorithms based on the KL expansion typically operate in discrete-time or require some sort of time-discretization, namely, sampling \cite{119724}. 
In contrast to other processes that are bandlimited or have a finite rate of innovation \cite{eldar2015sampling}, the \emph{self-similarity} property of the Wiener path implies that its fluctuations scale with time resolution. It is therefore impossible to obtain an equivalent discrete-time representation of the Wiener process by sampling its path \cite{vervaat1985sample}. Consequently, Berger's achievability scheme, as well as any source coding approach that is based on transforming the Wiener path to discrete coefficients, is prone to sampling error in addition to the distortion due to the bitrate constraint. 
\par 
\begin{figure}
\begin{center}
\begin{tikzpicture}[node distance=2cm,auto,>=latex]
  \node at (0,0) (source) {$W^T$};

  \node [coordinate, right of = source,node distance = 2.5cm] (smp_in) {};
  \node [coordinate, right of = smp_in,node distance = 1cm] (smp_out){};
  \node [coordinate,above of = smp_out,node distance = 0.6cm] (tip) {};
\fill  (smp_out) circle [radius=2pt];
\fill  (smp_in) circle [radius=3pt];
\fill  (tip) circle [radius=2pt];
\node[left,left of = tip, node distance = 0.8 cm] (ltop) {$f_s$};
\draw[->,dashed,line width = 1pt] (ltop) to [out=0,in=90] (smp_out.north);
\node at (7,0) (right_edge) {};
\node [below of = right_edge, node distance = 1.7cm] (right_b_edge) {};

\node [right] (dest) [below of=source, node distance = 1.7cm] {$\widehat{W}^T$};
\node [int] (dec) [right of=dest, node distance = 2cm] {$\mathrm{Dec.}$};
\node [int] (enc) [right of = dec, node distance = 3.9cm]{$\mathrm{Enc.}$};

  \draw[-,line width=2pt] (smp_out) -- node[above, xshift = 0.8cm] {$\bar{W}^{\lfloor Tf_s \rfloor}$  } (right_edge);
  \draw[-,line width = 2]  (right_edge.west) -| (right_b_edge.east);
    \draw[->,line width = 2]  (right_b_edge.east) -- (enc.east);
   \draw[->,line width=2pt] (enc) -- node[above] {$ TR $ bits} (dec);

   \draw[->,line width=2pt] (dec) -- (dest);
    \draw[->,line width=2pt] (source) -- (smp_in);
    \draw[line width=2pt]  (smp_in) -- (tip);
    \draw[dashed, line width = 1pt] (smp_in)+(-0.5,-0.3) -- +(-0.5,1) -- +(1.5,1) -- + (1.5,-0.3) --  node[below] {sampler} +(-0.5,-0.3);
    
\end{tikzpicture}

\caption{\label{fig:system_model}
Uniform sampling and source coding system model.
\vspace{-20pt}}
\end{center}
\end{figure}
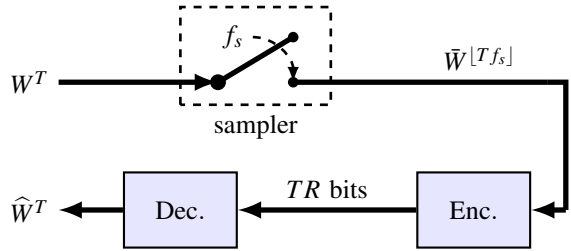
\par
In order to account for the effect of sampling on the overall distortion in encoding the Wiener process, we consider in this work a combined setup of sampling and source coding as described in Fig.~\ref{fig:system_model}. In this setup the continuous-time Wiener process $W_{(\cdot)}=\left\{W_t\right\}_{t\geq 0}$ is first uniformly sampled at rate $f_s$ over the time interval $[0,T]$, resulting in the finite dimensional random vector of samples $\bar{W}^{\lfloor T f_s \rfloor}$. This vector is then encoded using no more than $TR$ bits, and ultimately the original Wiener path is reconstructed from this encoded version under a mean squared error (MSE) distortion criterion. We analyze the minimal distortion in the asymptotic regime of a large time horizon $T$, as a function of the sampling rate $f_s$, and the \emph{bitrate} $R$. The optimal tradeoff among the three is described by the function $D(f_s,R)$, providing the minimal quadratic distortion in reconstructing the Wiener path when the sampling rate is $f_s$ and the bitrate is $R$. Consequently, the ratio beteween $D(f_s,R)$ and the DRF of the Wiener process at bitrate $R$ from \cite{berger1970information} represents the excess distortion due to a rate $f_s$ sampling constraint in encoding the Wiener process.

\begin{figure}
\begin{center}
\begin{tikzpicture}[node distance=3cm,auto,>=latex]
 \node at (-0.5,0) (source) {$\bar{W}^{\lfloor T f_s \rfloor }$};

\node[int1, fill = red!30, node distance = 1.75cm, align = center] at (1,1) (est) {\small $\mathrm{estimator}$ \\ {\small  (w.r.t. $W^T$)}}; 
\node[int1, right of = est, align = center, node distance = 3.3cm] (enc)
{\small $\mathrm{Enc.}$ \\  {\small (w.r.t. $\tilde{W}^T$)}}; 
\node[int1, right of = enc, align = center, node distance = 3cm] (dec)
{\small $\mathrm{Dec.}$ \\ {\small  (w.r.t. $\tilde{W}^T$)} }; 
\node[right of = dec, node distance = 1.5cm] (dest) {$\widehat{W}^T$};
\draw[->, line width = 1pt] (source) |- (est.west);
\draw[->, line width = 1pt] (est) -- node[above] {$\tilde{W}^T$} (enc);
\draw[->, line width = 1pt] (enc) -- node[above] {\small $TR$ bits} (dec);
\draw[->, line width = 1pt] (dec)-- (dest);

\node[int1, below of = est, align = center, node distance = 2cm] (enc1)
{\small $\mathrm{Enc.}$ \\ (w.r.t. $\bar{W}^{\lfloor T f_s \rfloor }$)}; 
\node[int1, below of = enc, align = center, node distance = 2cm] (dec1)
{\small $\mathrm{Dec.}$ \\ {\small (w.r.t. $\bar{W}^{\lfloor Tf_s \rfloor }$)}}; 
 \node[int1, fill = red!30, below of = dec, node distance = 2cm, align = center] (est1) {\small $\mathrm{estimator}$ \\ {\small  (w.r.t. $W^T$)} }; 

\node[right of = est1, node distance = 1.5cm] (dest1) {$\widehat{W}^T$};
\draw[->, line width = 1pt] (source) |- (enc1.west);
\draw[->, line width = 1pt] (est1) --  (dest1);
\draw[->, line width = 1pt] (enc1) -- node[above] {\small $TR$ bits} (dec1);
\draw[->, line width = 1pt] (dec1) -- node[above] {$\widehat{\bar{W}}^{\lfloor Tf_s  \rfloor}$} (est1);
\end{tikzpicture}

\caption{\label{fig:ec_vs_ce}
Two source coding approaches. Up: Estimate-and-Compress (EC). Down: Compress-and-estimate (CE). EC achieves the minimal distortion in the combined sampling and source coding problem of Fig.~\ref{fig:system_model}. CE is sub-optimal but does not require sampling rate information at the encoder.
}
\end{center}
\end{figure}
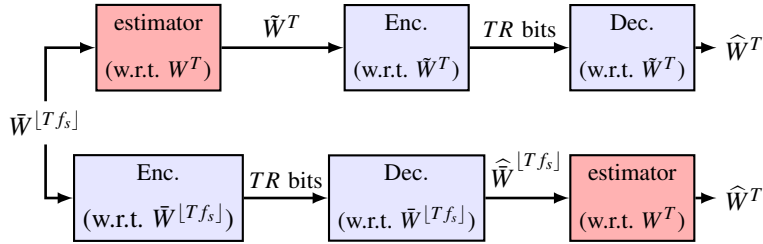

In the combined sampling and source coding setting illustrated in Fig.~\ref{fig:system_model}, the encoder has no direct access to the realizations of the source it is trying to describe. Therefore, this problem falls within the regime of \emph{indirect} source coding (ISC), a.k.a. \emph{remote} or \emph{noisy} source coding \cite[Ch 3.5]{berger1971rate}. It is well-known that the optimal source code in such problems is attained by first estimating the original path from the samples, and then encoding this estimate using a source code that is optimal with respect to the estimated process \cite{1054469}. As we explain in Section~\ref{sec:main_results}, this \emph{estimate-and-compress} (EC) strategy allows us to characterize $D(f_s,R)$ by considering two separate problems: A minimal MSE (MMSE) estimation problem that does not involve coding, and a standard source coding problem with respect to the process resulting from this estimation. \par
Although the EC strategy leads to the minimal distortion under a sampling rate $f_s$ and a bitrate $R$, it has an important caveat: it requires availability of the sampling rate $f_s$ at the encoder, since both the estimation and compression steps depend on $f_s$. It is easy to imagine a scenario where the encoder is either not informed of $f_s$ or is completely unaware of the continuous-time origin of the samples it is given. For instance, this situation arises when a model based on a continuous-time Wiener process is fitted to the measurements only after these were quantized or compressed to satisfy the bit constraints in the acquisition process. 
In this scenario of missing information on the sampling rate or the existence of an underlying continuous-time process, the encoder and decoder may operate according to a \emph{compress-and-estimate} (CE) source coding strategy, as illustrated in Fig.~\ref{fig:ec_vs_ce}: the encoder employs an optimal source code to compress the discrete-time samples subject to a quadratic distortion criterion; the decoder, provided with the sampling rate, estimates the continuous-time path from the output of the encoder. Clearly, the resulting distortion under a CE approach provides an upper bound on $D(f_s,R)$, and describes the excess distortion in lieu of sampling rate information at the encoder or the price of ignoring the continuous-time nature of the samples. 
 \par
%
In this work we analyze the distortion under a CE source coding approach when encoding is performed with respect to the random codebook that attains the DRF of the discrete-time vector of samples. We show that although encoding with respect to the samples as in CE is not equivalent to encoding with respect to the estimation of the Wiener process from its samples, the ratio between the distortion under this CE coding scheme is not more than $1.027$ times higher than $D(f_s,R)$. That is, while information removed at the encoding stage differs between CE and EC, the performance difference between the two is relatively minor. 

\subsection{Contributions}
The main contribution of this paper is the characterization of the expected MSE distortion in the following cases:
\begin{itemize}
\item[(1)] Minimal distortion under all possible bitrate $R$ representations of uniform rate $f_s$ samples of the Wiener process (the function $D(f_s,R)$). 
\item[(2)] Minimal distortion in the CE scenario where the uniform rate $f_s$ samples are encoded using a random codebook chosen to minimize the MSE distortion with respect to the sequence of samples. We denote this distortion by $D_{CE}(f_s,R)$. 
\end{itemize}
The characterization of $D(f_s,R)$ is achieved by first providing an information theoretic description of this function as the solution of a sequence of optimization problems involving only probability density functions of limited mutual information rate. This characterization leads to a similar information expression for the process obtained by estimating the Wiener process from its uniform samples. The KL transform of this signal defines a sequence of finite-rank operators, and the expression for $D(f_s,R)$ is given in terms of the limiting eigenvalue distribution of these operators. Finally, we analyze the ratio between $D(f_s,R)$ and the DRF of the Wiener process from \cite{berger1970information}, as well as the the ratio between $D(f_s,R)$ and the MMSE in estimating the Wiener process from its rate $f_s$ uniform samples. These ratios describe the excess distortion due to sampling in the lossy compression of the Wiener process, and the excess distortion due to a bitrate constraint in the sampling of the Wiener process, respectively. We show that both excess distortions are only a function of the number of bits per sample $R/f_s$. \par
As opposed to $D(f_s,R)$ that describes the minimal distortion under any source code, in CE we consider a specific source code: the compression of the samples of the Wiener process using the achievable scheme for the DRF of the discrete-time Wiener process described in \cite{berger1970information}. We show that when this source coding scheme is employed with coding bitrates converging to $R$ from above as the time horizon goes to infinity, the resulting distortion converges to an expression we denote as $D_{\CE}(f_s,R)$. This expression is defined in terms of the asymptotic eigenvalue distribution of the operators defining $D(f_s,R)$, as well as the asymptotic eigenvalues distribution of the covariance matrix of the samples of the Wiener process. Finally, we compare $D_{\CE}(f_s,R)$ with $D(f_s,R)$, and conclude that the ratio between the two is bounded from above by $1.027$. That is, the performance loss in using CE compared to the optimal source coding scheme is at most $2.7\%$. This loss can be seen as the penalty in ignoring the continuous-time origin of the samples at the encoder, or not knowing the sampling rate at which these samples were obtained.


\subsection{Related Works}
The DRF of the Wiener process was derived by Berger in \cite{berger1970information}. Gray established similar results for the more general family of unstable auto-regressive processes \cite{1054470}. See also \cite{sahai2006necessity} for applications of Berger's and Gray's works in reliable communication of unstable processes \cite{sahai2006necessity}. 
The work of \cite{neuhoff2013information} derives conditions under which the DRF of a continuous-time stationary Gaussian process, possibly non-bandlimited, can be attained by encoding its samples obtained at asymptotically high rates. In contrast to \cite{neuhoff2013information}, here we consider a non-stationary source signal and are interested in the optimal source coding performance under a fixed sampling rate, rather than the distortion in the limit of infinitesimally dense sampling grids. %
Closely related works are \cite{1184140, xiong2004distributed}, which consider reconstruction from sampling using sensor networks followed by distributed lossy compression. The work \cite{7039920} considers a problem of sampling a multi-dimensional Wiener process with limited resources, although without a constraint on the number of bits in representing the samples. 
%
A specific scenario of sampling the Wiener process under a communication constraint is considered in \cite{DBLP:journals/corr/SunPU17}, to which our function $D(f_s,R)$ provides performance lower bounds. 
Unlike the results described above, the combined sampling and source coding setting of Fig.~\ref{fig:system_model} 
allows us to derive the optimal tradeoff between distortion, bitrate, and sampling rate under any digital representation of these samples. We explored such a tradeoff in \cite{Kipnis2014, KipnisBitrate, kipnisSPM2018} for second-order Gaussian stationary processes. 
\par
It is well known that in ISC settings such as in Fig.~\ref{fig:system_model}, the minimal distortion is attained via an EC strategy
 \cite[Ch. 3.5]{berger1971rate}, \cite{1054469}, \cite{1056251}. 
The CE setting of \cite{KipnisRini2017} was proposed in order to study the performance in cases where estimation before compression is impossible due to lack of computation resources, missing information for performing this estimation such as the sampling rate in our setting, or simply an ad-hoc system design that is unaware of the indirect source. CE performance was recently explored in a compressed-sensing framework when the sampling matrix is unavailable at the encoder \cite{KipnisCS}.  
In contrast to the examples in \cite{KipnisRini2017}  and the ISC setting of \cite{KipnisCS}, here the relation between the source signal and its observations at the encoder is deterministic.
 \\

The rest of this paper is organized as follows: In Section~\ref{sec:problem_formulation}, we define a combined sampling and source coding problem for the Wiener process. In Section~\ref{sec:preliminary} we provide preliminary results that are based on known source coding results with respect to the Wiener process. In Section~\ref{sec:main_results} we characterize the minimal distortion in the combined sampling and source coding problem of Fig.~\ref{fig:system_model}. In Section \ref{sec:ce}, we consider the distortion under the CE approach. Concluding remarks are provided in Section~\ref{sec:conclusion}.

\section{Problem Formulation \label{sec:problem_formulation}}
Let $W_{(\cdot)} = \left\{W_t,\,t  \geq 0 \right\}$ be a continuous-time Gaussian process with zero mean, autocovariance function
\[
K_W(t,s) \triangleq \mathbb E \left[W_tW_s \right] = \sigma^2 \min\{t,s\},\quad t,s \geq 0,
\]
and $W_0 = 0$ almost surely. The standard definition of the Wiener process also requires that each realization of $W_{(\cdot)}$ has almost surely continuous paths \cite{karatzas1991brownian}. In our setting, however, only the weaker assumption of almost surely Riemann integrability of the paths is required so that this path can be approximated in discrete-time in the $\Ltwo$ sense. \par
We consider the system depicted in Fig.~\ref{fig:system_model} to describe the random waveform $W^T\triangleq \left\{W_t,\,t\in[0,T] \right\}$ using a code of rate $R$ bits per unit time. Unlike in the regular source coding problem for the Wiener process considered in \cite{berger1970information}, we assume that $W^T$ is first uniformly sampled at frequency $f_s$. Set 
\[
N_T \triangleq \lfloor T f_s \rfloor
\]
to be the number of samples obtained by sampling the Wiener process over $[0,T]$, and denote the vector of samples by
\begin{equation}
\bar{W}^{N_T} \triangleq \left\{ W_{n/f_s},\,n\in \mathbb N \cap [0,f_sT]  \right\}. \label{eq:sampling_relation}
\end{equation}
Next, an \emph{encoder} 
\[
f: \mathbb R^{N_T} \rightarrow \left\{1,\ldots,2^{ \lfloor T R \rfloor} \right\}
\]
maps $\bar{W}^{N_T}$ to an index out of $2^{\lfloor T R  \rfloor}$ possible indices. The \emph{decoder}, upon receiving the index $f(\bar{W}^{N_T})$, provides a \emph{reconstructed} waveform $\widehat{W}^T = \left\{\hat{W}_t,\,t \in [0,T] \right\}$.
\par 
The optimal performance theoretically achievable (OPTA) in terms of the distortion in estimating $W_{(\cdot)}$ from its samples $\bar{W}_{[\cdot]}$ is defined as 
\begin{equation}
\label{eq:D_def}
D(f_s,R) = \liminf_{T \to \infty} D^{\OPTA}_T (R)
\end{equation}
where
\[
D^{\OPTA}_T (R) = \inf_{\enc-\dec} \frac{1}{T} \int_0^T \mathbb E\left(W_t-\widehat{W}_t \right)^2 dt,
\]
and the infimum is taken over all encoders and decoders to and from a set of at most $TR$ elements. 
We note that since $W_0 =0$, replacing the limit inferior in \eqref{eq:D_def} with infimum over $T$ leads to the trivial solution $D(f_s,R)=0$. Our definition of OPTA avoids this degenerate case. 
\par
Without loss of generality, the OPTA can be written as 
\begin{equation}
\label{eq:OPTA_def}
D(f_s,R) =\liminf_{T\rightarrow \infty} ~\inf_{f} ~ \mmse\left(W^T | f(\bar{W}^{N_T}) \right),
\end{equation}
where 
\[
\mmse\left(W^T | f(\bar{W}^{N_T}) \right) \triangleq \int_0^T \mathbb E \left( W_t - \mathbb E \left[W_t |f(\bar{W}^N_T)  \right] \right)^2 dt
\]
is the MMSE in estimating $W^T$ from $f(\bar{W}^{N_T})$. That is, compared to the definition of $D(f_s,R)$ in \eqref{eq:OPTA_def}, we eliminate the dependency on the decoder by assuming that this provides the MMSE estimate of $W^T$ given the output of the encoder. \\

The main goal of this paper is to derive an expression for $D(f_s,R)$ in closed form, as well as to characterize $\mmse(W^T | f(\bar{W}^{N_T})$ under an encoder $f$ that follows the CE approach. 
Before doing so, we explore, in the next section, the connection between these two distortions to the DRF of the Wiener processes derived in \cite{berger1970information} without sampling, and to the MMSE in sampling the Wiener process without a bitrate constraint. 

\section{Preliminaries}
 \label{sec:preliminary}
In this section we review known results on the optimal MSE attainable in encoding the continuous and discrete-time Wiener processes, and derive connections between these results and the combined sampling and source coding problem of Fig.~\ref{fig:system_model}. In particular, we show how these results lead to upper and lower bounds to $D(f_s,R)$. The notation and preliminary results provided in this section are used throughout the paper.\\

\subsection{The Distortion-Rate Function of the Wiener Process}

It is shown in \cite{berger1970information} that the OPTA in encoding the Wiener process are given by Shannon's DRF of this process:
\begin{equation} \label{eq:DR_BM_cont}
 D_W(R) = \frac{2\sigma^2}{\pi^2 \ln 2} R^{-1} \approx 0.292 \sigma^2 R^{-1}.
\end{equation}
That is, the minimal expected distortion attainable in recovering a Wiener path from its encoded version is inversely proportional to the number of bits per unit time in this encoding. 
Compared to the combined sampling and source coding problem described in Fig.~\ref{fig:system_model} where the source code is constrained to be a function of its uniform samples, \eqref{eq:DR_BM_cont} represents the OPTA when the source code is any functional of the Wiener path. \par
The achievability of \eqref{eq:DR_BM_cont} is based on the following procedure: divide the interval $[0,T]$ into $L$ identical sub-intervals, each of length $T' = T/L$. For each $l=0,\ldots L-1$, expand the $l$th section of path $W^T$ according the the KL expansion of the Wiener process over the interval $[0,T']$. The $k$th KL coefficient in this expansion is given by
\begin{equation} \label{eq:KL_Wiener}
 \frac{1}{T'} \int_0^{T'} \phi_k(t) \left( W_{t-lT'} -W_{lT'}\right) dt, \quad k\in \mathbb N,
\end{equation}
where $\phi_k(t)$ is the $k$th KL eigenfunction of the Fredholm integral equation of the second kind \cite{zemyan2012classical} over the interval $[0,T']$ with Kernel $K_W(t,s)$. The coefficients \eqref{eq:KL_Wiener} constitute a set of $L$ independent sequences of i.i.d.\ Gaussian random variables. 
Each of these sequences is encoded using a single code of $2^{R_lT}$ codewords that is optimal with respect to the scalar Gaussian distribution with variance equal to the $l$th KL eigenvalue, where $R_l$ is determined using Kolmogorov's waterfilling formula \cite{1056823}. The reconstruction waveform is obtained by using the decoded KL coefficients and the KL eigenfunctions. Finally, in order to avoid unbounded distortion due to inaccurate block starting locations in reconstruction, Berger suggested to encode the sequence of block starting locations $\left\{W_{lT'},\,l=0,\ldots,L-1 \right\}$ using a separate bitstream; he showed that a delta modulator can achieve arbitrary precision in encoding these locations without increasing the bitrate.
\par
It was shown in \cite[Sec. IV]{berger1970information} 
that by taking $T$ and $L$ to infinity such that $L/T$ goes to zero,  the bitrate required to encode $\left\{W_{lT'},\,l=0,\ldots,L-1 \right\}$ is negligible, hence it can be provided to the decoder without increasing the overall bitrate. As a result, the scheme above attains distortion as close to $D_W(R)$ as desired while keeping the bitrate at most $R$.  \par
However, as explained in the introduction, it is extremely challenging to realize the integration in \eqref{eq:KL_Wiener} 
without sampling first the analog Wiener path. Instead, here we consider source coding schemes for $W_{(\cdot)}$ which assume that only the samples are available at the encoder, rather than the entire continuous-time path. \par
The samples of $W_{(\cdot)}$ define a discrete-time Wiener process, and in what follows we consider the optimal performance in encoding this process according to an MSE criterion subject to a bitrate constraint. 

\subsection{The Distortion-Rate Function of the Discrete-time Wiener Process}

The autocovariance function of the discrete-time process $\bar{W}_{[\cdot]}= \left\{W_{n/f_s},\,n=0,1,\ldots \right\}$ obtained by sampling $W_{(\cdot)}$ at rate $f_s$, is given by
\begin{align*}
\mathbb E \left( \bar{W}_n \bar{W}_k \right) & = \mathbb E \left( W_{n/f_s} W_{k/f_s} \right) = \frac{\sigma^2}{f_s} \min\left\{ n,k\right\}. 
\end{align*}
The process $\bar{W}_{[\cdot]}$ is called a discrete-time Wiener process with intensity $\sigma^2/f_s$ (a.k.a. a Gaussian random walk). A closed form expression for its DRF was also derived in \cite{berger1970information}, and can be written as follows:
\begin{equation} \label{eq:DR_BM_disc}
\begin{split}
D(\bar{R}_\theta) & =  \frac{\sigma^2}{f_s} \intzerotoone \min \left\{ S_{\bar{W}}(\phi),\theta \right\} d\phi, \\
\bar{R}_\theta & = \frac{1}{2} \intzerotoone \log^+ \left[ S_{\bar{W}}(\phi)/\theta \right] d\phi,
\end{split}
\end{equation}
where $\bar{R}$ is the amount of bits per sample\footnote{
These units of measurements are consistent with our previous notations: the DRF of a source is evaluated as the number of bits per source symbol available for the code.} of the code
and
\begin{equation} \label{eq:S_bar}
S_{\bar{W}}(\phi) \triangleq  \frac{1}{4 \sin^2 \left( \pi \phi /2  \right) }
\end{equation}
is the asymptotic density of the eigenvalues of the matrix with entries $\min\{n,k\}$, $n,k=0,\ldots,N-1$, as $N$ goes to infinity. 
Expression \eqref{eq:S_bar} gives the distortion as a function of the rate, or the rate as a function of the distortion, through a joint dependency on the parameter $\theta$. Such a parametric representation is said to be of a \emph{waterfilling} form, since only the part of $S_{\bar{W}}(\phi)$ below the water level parameter $\theta$ contributes to the distortion.
\par
Keeping the bitrate $R=f_s \bar{R}$ fixed and increasing $f_s$, we see that the asymptotic behavior of the DRF of $\bar{W}_{[\cdot]}$ as $f_s$ goes to infinity is given by \eqref{eq:DR_BM_disc} when $\bar{R}$ goes to zero or, equivalently, when $\theta$ goes to infinity. The latter can be obtained by expanding both expressions in \eqref{eq:DR_BM_disc} according to $\theta^{-1}$, which, after eliminating $\theta$ leads to 
\begin{equation} \label{eq:dr_asym}
\begin{split}
 D_{\bar{W}}(\bar{R}) & \sim \frac{\sigma^2}{f_s} \left( \frac{2}{\pi^2 \ln 2 \bar{R}} + \frac{\bar{R} \ln 2}{12} + O(\bar{R}^{-2})  \right) \\
  & = \frac{2\sigma^2}{\pi^2 \ln2} R^{-1} + \frac{ \sigma^2 \ln 2}{12 }\frac{R}{f_s^2} + O(f_s^{-3}).
 \end{split}
\end{equation}
Note that the first term in \eqref{eq:dr_asym} is the DRF of the continuous-time Wiener process \eqref{eq:DR_BM_cont}. Thus, we have proven the following:
\begin{prop} \label{prop:fs_to_infinity}
Let $\bar{W}_{[\cdot]}$ be the process obtained by uniformly sampling the Wiener process $W_{(\cdot)}$ at sampling rate $f_s$. Then
\[
 \lim_{f_s \rightarrow \infty} D_{\bar{W}}(R/f_s) =  D_{W}(R).
\] 
\end{prop}
In fact, $D_{\bar{W}}(R/f_s)$ is monotonically increasing in $f_s$ so that 
\begin{equation} \label{eq:discrete_to_cont_sup}
 \sup_{f_s > 0} D_{\bar{W}}(R/f_s) =  D_{W}(R).
\end{equation}
Proposition~\ref{prop:fs_to_infinity} provides an intuitive explanation for a fact observed in \cite{berger1970information}: the DRF of a discrete-time Wiener process at high distortion behaves as the DRF of a continuous-time Wiener process. Proposition \ref{prop:fs_to_infinity} shows that this fact is simply the result of evaluating the DRF of the discrete-time Wiener process $\bar{W}_{[\cdot]}$ at high sampling rates, while holding the bitrate $R$ fixed. Due to the high sampling rate, the number of bits per sample $\bar{R}=R/f_s$ goes to zero and the DRF of the discrete-time Wiener process is evaluated at the large distortion (low bit) limit. The fact that $D_{\bar{W}}(R/f_s)$ is monotonically increasing in $f_s$ implies that the path of the sampled Wiener process becomes harder to describe as the frequency at which those samples are obtained increases. \par
%
Since the paths of the Wiener process are Riemann integrable, the $\Ltwo$ distance between any reasonable reconstruction technique (e.g., linear interpolation) of $W^T$ from $\bar{W}^{N_T}$ converges to zero as $f_s$ goes to infinity. Therefore, in addition to convergence of their respective DRFs as expressed in Proposition~\ref{prop:fs_to_infinity}, the path of the optimal reconstruction of $\bar{W}^{N_T}$ from its encoded version converges to the path of the reconstruction of $W^T$ from its encoded version in the $\mathrm L_2$ sense. It follows that a distortion arbitrarily close to $D_W(R)$ can be obtained by the following procedure: 
\begin{itemize}
\item[ (i) ] Choose $T$ large enough such that the distortion under Berger's achievability scheme for $W^T$ is close to $D_W(R)$. 
\item[(ii)\,] Take $f_s$ large enough such that $D_W(R)$ is close to $D_{\bar{W}}(R/f_s)$. 
\item[(iii)] Encode $\bar{W}^{N_T}$ using a code that attains distortion close to $D_{\bar{W}}(R/f_s)$. 
\item[(iv)] Estimate $W^T$ from the encoded version of $\bar{W}^{N_T}$.
\end{itemize}
 Since this procedure falls under the system of Fig.~\ref{fig:system_model}, we necessarily have $\liminf_{f_s \to \infty} D(f_s,R) \leq D_W(R)$, and hence
\begin{equation} \label{eq:lim_fs}
\lim_{f_s \to \infty } D(f_s,R) = D_W(R). 
\end{equation}
Following the characterization of $D(f_s,R)$ in Section~\ref{sec:main_results} below, we show that the convergence in \eqref{eq:lim_fs} is inversely quadratic in $f_s$. 

%


We now consider the other extreme in the combined sampling and source coding of Fig.~\ref{fig:system_model}: finite sampling rate and infinite bitrate.

\subsection{Minimal MSE under Sampling \label{subsec:mmse}}
\begin{figure}
\begin{center}
\begin{tikzpicture}
\node at (0,0) {\includegraphics[scale = 0.4]{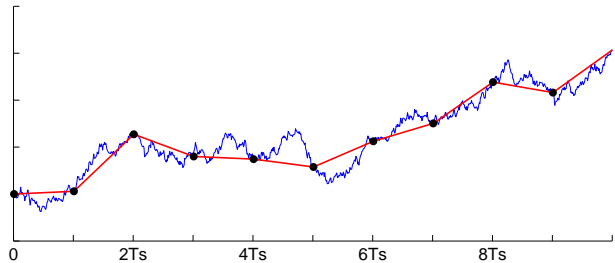}};
\fill[white] (3.6,-1.5) rectangle (4.2,-1.8);
\end{tikzpicture}
\caption{\label{fig:sample_path} Sample paths of  $W_{(\cdot)}$ (blue), $\bar{W}_{[\cdot]}$ (black dots) and $\tilde{W}_{(\cdot)}$ (red). }
\end{center}
\end{figure}

A trivial lower bound on $D(f_s,R)$ is obtained by relaxing the bitrate constraint in Fig.~\ref{fig:system_model} by letting $R \rightarrow \infty$. Under this relaxation, the function $D(f_s,R)$ reduces to the MMSE in estimating the Wiener process from its samples, denoted as $\mmse\left(W_{(\cdot)}|\bar{W}_{[\cdot]} \right)$. For $t>0$, denote by $t^+$ and $t^-$ the two points on the grid $\mathbb Z T_s$ closest to $t$, namely, $t^- = \lfloor t/T_s \rfloor T_s$ and $t^+ = \lceil t/T_s \rceil T_s$. Because of the Markov property of $W_{(\cdot)}$, the MMSE in estimating $W_{t}$ from the process $\bar{W}_{[\cdot]}$ is given by linear interpolation between these two points:
\begin{align}
\tilde{W}_t \triangleq & \mathbb E \left[W_t| \bar{W} _{[\cdot]}\right] = \mathbb E \left[W_t| W_{t^+},W_{t^-} \right] \nonumber \\ 
& =  \frac{t^+-t}{T_s}W_{t^-} + \frac{t-t^-}{T_s}W_{t^+}, \label{eq:W_tilde_def}
\end{align}
where $T_s = 1/f_s$. See Fig.~\ref{fig:sample_path} for an illustration of  the path of the processes $W_{(\cdot)}$, $\bar{W}_{[\cdot]}$ and $\tilde{W}_{(\cdot)}$.

The instantaneous estimation error $B_t \triangleq W_t-\tilde{W}_t$ defines a \emph{Brownian bridge} on any interval whose endpoints are on the grid $\mathbb Z T_s$. The autocovariance function of $B_{(\cdot)}$ is given by
\begin{align}
& K_B(t,s)  = \mathbb E \left[ B_tB_s \right] \label{eq:cov_brownian_bridge} \\
& = \frac{\sigma^2}{T_s} \begin{cases}  (t^+ -t\vee s)(t\wedge s-t^- ) & nT_s \leq  t,s \leq (n+1)T_s, \\
0 & \mathsf{otherwise},   \end{cases} \nonumber
\end{align}
where $t\vee s$ and $t \wedge s$ denote the maximum and minimum of $\{t,s\}$, respectively. We conclude that 
\[
\mmse\left({W_t|\bar{W}_{[\cdot]}} \right)=\mmse\left(W_t|\bar{W}_{N_T}, \bar{W}_{N_T+1} \right)= K_B(t,t),
\]
and the average MMSE in estimating $W_{(\cdot)}$ from $\bar{W}_{[\cdot]}$ equals
\begin{align}
\mmse\left(W_{(\cdot)}|\bar{W}_{[\cdot]} \right) & = \lim_{T\rightarrow \infty} \frac{1}{T} \int_0^T \mmse\left({W_t|\bar{W}_{[\cdot]}} \right) dt \label{eq:mmse_value}
\\ \nonumber
& = \frac{1}{T_s} \int_{nT_s}^{(n+1)T_s} K_B(t,t) dt = \frac{\sigma^2 T_s}{6} = \frac{\sigma^2}{6f_s}. 
\end{align} 
For future use, we introduce the notation
\[
\mmse(f_s) \triangleq \mmse\left(W_{(\cdot)}|\bar{W}_{[\cdot]} \right)= \frac{\sigma^2}{6f_s}. 
\]

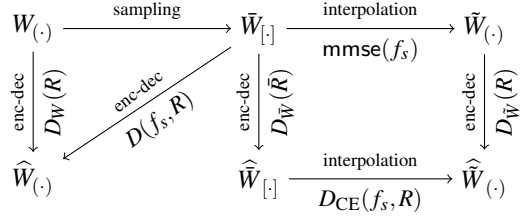
\begin{figure}
\begin{center}
\begin{tikzpicture}
\node (source) at (0,0) {$W_{(\cdot)}$};
\node[right of = source, node distance = 3cm] (samples)  {$\bar{W}_{[\cdot]}$};
\node[right of = samples, node distance = 3cm] (interp)  {$\tilde{W}_{(\cdot)}$}; 
\node[below of = source, node distance = 2cm] (source-est) {$\widehat{W}_{(\cdot)}$};
\node[below of = samples, node distance = 2cm] (samples-est) {$\widehat{\bar{W}}_{[\cdot]}$};
\node[below of = interp, node distance = 2cm] (interp-est) {$\widehat{\tilde{W}}_{(\cdot)}$};
\draw[->] (source)  -- node[above, rotate=90] {\scriptsize enc-dec} node[below, rotate=90] {\small $D_W(R)$} (source-est);
\draw[->] (samples)  -- node[above, rotate=90] {\scriptsize enc-dec} node[below, rotate=90] {\small $D_{\bar{W}}(\bar{R})$} (samples-est);
\draw[->] (interp)  -- node[above, rotate=90] {\scriptsize enc-dec} node[below, rotate=90] {\small $D_{\tilde{W}}(R)$} (interp-est);
\draw[->] (samples)  -- node[above, rotate = 37] {\scriptsize enc-dec} node[below, rotate = 37] {\small $D(f_s,R)$} (source-est);

\draw[->] (source)  -- node[above] {\scriptsize sampling} (samples);

\draw[->] (samples)  -- node[above] {\scriptsize interpolation} node[below] {\small $\mmse(f_s)$} (interp);

\draw[->] (samples-est)  -- node[above, align = center] {\scriptsize  interpolation} node[below, align = center] {\small $D_{\CE}(f_s,R)$} (interp-est);


\end{tikzpicture}
\caption{\label{fig:cascade_description} Relations among the processes $W_{(\cdot)}$, $\bar{W}_{[\cdot]}$, $\tilde{W}_{(\cdot)}$ and their associated distortion functions. Each reconstruction operation is associated with a distortion quantity. In this paper we show that $D_{\tilde{W}}(R)\leq D_{\bar{W}}(R/f_s) \leq D_W(R) \leq D(f_s,R) \leq D_{\CE}(f_s,R) \leq \mmse(f_s) + D_{\bar{W}}(R/f_s)$, where $D(f_s,R) = \mmse(f_s) + D_{\tilde{W}}(R)$. }
\end{center}
\end{figure}

From properties of the optimal MSE estimator, it follows that for any $T>0$,
\begin{equation} 
\mmse\left(W^T | f(\bar{W}^{N_T}) \right) = \mmse\left(W^T| \bar{W} \right) + \mmse \left( \tilde{W}^T | f(\bar{W}^{N_T} )\right).
\label{eq:separation_operational_T}
\end{equation}
Since the optimization in \eqref{eq:OPTA_def} is only over the term $\mmse \left( \tilde{W}^T | f(\bar{W}^{N_T} )\right)$,  \eqref{eq:separation_operational_T} implies that encoding $\bar{W}^{N_T}$ to best describe $W^T$ is equivalent to encoding $\bar{W}^{N_T}$ to best describe $\tilde{W}^T$. As a result, we conclude that the optimal encoding strategy for the system in Fig.~\ref{fig:system_model} is \emph{estimate-and-compress}: The encoder first estimates $W^T$ from the samples $\bar{W}^{N_T}$, and then applies an optimal source code to compress the estimate $\tilde{W}^T$ subject to the bitrate constraint. Furthermore, it follows that the OPTA can be written as
\begin{equation}
\label{eq:separation}
D(f_s,R) = \mmse(f_s) + D_{\tilde{W}}(R), 
\end{equation}
where $D_{\tilde{W}}(R)$ is the OPTA in encoding the continuous-time process $\tilde{W}_{(\cdot)}$ of \eqref{eq:W_tilde_def} at rate $R$. In other words, \eqref{eq:separation} reduces the problem of deriving $D(f_s,R)$ to that of deriving the OPTA in encoding at rate $R$ the MMSE estimation of $W_{(\cdot)}$ from its uniform rate $f_s$ samples. This decomposition of the OPTA can be seen as a special case of a more general result dicussed in \cite{1054469}. 
The relationships among the various processes and the distortion functions introduced thus far are illustrated in the diagram of Fig.~\ref{fig:cascade_description}. \\

Before considering the OPTA with respect to $\tilde{W}_{(\cdot)}$, which we defer to Section~\ref{sec:main_results}, we explore the relation between $D(f_s,R)$ to the distortion in estimating the samples $\tilde{W}_{[\cdot]}$ from an arbitrary finite bit representation of these samples. This relation provides a first upper estimate for $D(f_s,R)$, and will be used in Section~\ref{sec:ce} below to characterize the performance under the CE approach. \par

\subsection{MSE in Discrete- and Continuous-Time}
Consider an arbitrary finite bit representation 
$f(\bar{W}^{N_T}) \in \{1,\ldots,2^{\lfloor TR \rfloor} \}$ of the samples $\bar{W}^{N_T}$ in the system of Fig.~\ref{fig:system_model}. The two distortion functions associated with this representation are (1) the MMSE in estimating $W^T$, and (2) the MMSE in estimating $\bar{W}^{N_T}$. The following lemma connects these two distortions, and will be particularly useful in characterizing the distortion under CE in Section~\ref{sec:ce} below. 
\begin{lem} \label{lem:bounds}
Fix $T$, $f_s$, $R$, and any encoder $f : \mathbb R^{N_T} \rightarrow \left\{1,\ldots, 2^{N_T R} \right\}$. Let $\Delta_n \triangleq \bar{W}_n - \mathbb E \left[\bar{W}_n | f( \bar{W}^{N_T}) \right]$. 
The minimal MSE in estimating $W^T$ from $f( \bar{W}^{N_T})$ satisfies 
\begin{align}
& \mmse \left( W^T | f( \bar{W}^{N_T}) \right)  \geq \mmse(W^T|\bar{W}^{N_T})  \label{eq:lemma_exact_lower}  \\
&  \quad + \frac{2}{3}  \frac{1}{N_T} \sum_{n=0}^{N_T-1} \mathbb E \Delta_n ^2 + \frac{1}{3} \frac{1}{N_T}\sum_{n=0}^{N_T} \mathbb E \Delta_n \Delta_{n+1}, \nonumber
\end{align}
and 
\begin{align}
& \mmse \left( W^T | f( \bar{W}^{N_T}) \right) \leq \mmse(W^T|\bar{W}^{N_T})  \nonumber  \\ 
\quad  & + \frac{2}{3} \frac{1}{N_T+1} \sum_{n=1}^{N_T+1}  \mathbb E \Delta_n^2  +  \frac{1}{3(N_T+1)} \sum_{n=1}^{N_T} \mathbb E\Delta_n \Delta_{n+1}.
 \label{eq:lemma_exact_upper} 
\end{align}
\end{lem}
\begin{IEEEproof}
See Appendix~\ref{app:proofs}
\end{IEEEproof}
Lemma~\ref{lem:bounds} shows that for any finite bit representation of the samples, the MSE in recovering the samples and the MSE in recovering the continuous-time Wiener process cannot be too far from each other. 
%
%
An interesting corollary of Lemma~\ref{lem:bounds} arises if we consider a sequence of encoders $\{\bar{f}_N \}_{N\in\mathbb N}$ such that, together with the optimal MSE estimation of $\bar{W}^{N_T}$ from the output of the encoders, define a 
 \emph{good rate-distortion code} for $\bar{W}_{[\cdot]}$ \cite{kanlis1996typicality, weissman2005empirical}. The term good rate-distortion code refers to the fact that the distortion attained by encoding and decoding approaches the DRF of $\bar{W}_{[\cdot]}$, namely
\begin{equation}
\label{eq:opt_enc}
\lim_{N \rightarrow \infty} \mmse \left(\bar{W}_{[\cdot]} | \bar{f}_N(\bar{W}^{N}) \right) = D_{\bar{W}}( R / f_s).
\end{equation}
The existence of such a sequence follows from the source coding theorem with respect to $\bar{W}_{[\cdot]}$ proved in \cite{berger1970information}, and leads to the following upper bound on $D(f_s,R)$:
\begin{cor} \label{cor:upper_bound}
For any $R>0$ and $f_s>0$, let 
\begin{align}
D^U(f_s,R_\theta) & = \mmse(f_s) + D_{\bar{W}}(R_\theta/f_s) \nonumber \\
& = \frac{\sigma^2}{6f_s} + \frac{\sigma^2}{f_s} \int_0^1 \min \left\{ S_{\bar{W}}(\phi),\theta \right\}d \phi, \label{eq:upper_bound}
\end{align}
where $S_{\bar{W}}(\phi)$ is given in \eqref{eq:S_bar} and
\[
R_\theta = \frac{f_s}{2} \int_0^1 \log^+ \left[ S_{\bar{W}}(\phi)/\theta \right] d\phi.
\]
The OPTA in estimating a path of the Wiener process from any rate-$R$ encoding of its uniform samples at rate $f_s$ satisfies
\[
D(f_s,R) \leq D^U(f_s,R).
\]
\end{cor}

\begin{IEEEproof} See Appendix~\ref{app:proofs}
\end{IEEEproof}

By considering $D^U(f_s,R)$ in the two extreme cases of the ratio between bitrate to sampling rate we obtain estimates for the convergence rates of $D(f_s,R)$ to $D_W(R)$ and $\mmse(f_s)$ in the large sampling rate and bitrate asymptotic, respectively:
\subsubsection*{Low sampling rate} when $R \geq f_s$, \eqref{eq:upper_bound} reduces to 
\begin{equation} \label{eq:five_sixths_bound_special}
D^U(f_s,R) = \mmse(f_s) +  \frac{\sigma^2}{f_s}   2^{-2R/f_s}.
\end{equation}
In this regime we have
\begin{align*}
D(f_s,R) - \mmse(f_s) \leq D^U(f_s,R) - \mmse(f_s)
= \frac{\sigma^2}{f_s}  2^{-2R/f_s}.
\end{align*}
In particular, we conclude that for any $f_s>0$, 
\begin{equation}
\label{eq:R_to_infty}
\lim_{R\to \infty} D(f_s,R) = D_W(R). 
\end{equation}
\subsubsection*{High sampling rate} when $f_s$ is high compared to $R$, \eqref{eq:dr_asym} implies 
\begin{equation}
D^U(f_s,R) = \frac{\sigma^2}{6f_s} +  \frac{2\sigma^2} {\pi^2 \ln 2} R^{-1} + O(f_s^{-2}),
\end{equation}
and therefore
\[
D(f_s,R) - D_W(R) \leq D^U(f_s,R) - D_W(R) = O(f_s^{-1}). 
\]
In Section~\ref{sec:main_results} we will see that the upper bound $D^U(f_s,R)$ is loose except in trivial cases, and in particular that $D(f_s,R) - D_W(R) = O(f_s^{-2})$. Furthermore, in Section~\ref{sec:ce} we derive a closed form expression for the distortion attained by a particular good sequence of encoders with respect to $\bar{W}_{[\cdot]}$. This distortion is shown to be strictly smaller than $D^U(f_s,R)$ and strictly larger than $D(f_s,R)$.\\

So far we considered elementary properties of $D(f_s,R)$, and concluded that $D(f_s,R)$ is bounded from below by $\mmse(f_s) = \sigma^2/(6f_s)$ and by $D_W(R)$ of \eqref{eq:DR_BM_cont}, and from above by $D^U(f_s,R)$ of \eqref{eq:upper_bound}. We also showed that $D(f_s,R)$ converges to these expressions as $f_s$ or $R$ go to infinity, respectively. In the next section we provide an information theoretic characterization of $D(f_s,R)$, and use this characterization to derive it in closed form. 

\section{The Fundamental Distortion-Sampling-Bitrate Limit \label{sec:main_results}}
We now derive a closed form expression for the function $D(f_s,R)$ that describes the OPTA in recovering the Wiener process from an encoded version of its samples. This derivation is obtained by first proving a source coding theorem for the combined sampling and source coding problem of Fig.~\ref{fig:system_model}, and then evaluating the information expression resulting from this theorem. 

\subsection{A Combined Sampling and Source Coding Theorem \label{sec:source_coding}}

The OPTA in the combined sampling and source coding setting of Fig.~\ref{fig:system_model} is given by the following theorem.
\begin{thm} \label{thm:coding_theorem}
Define
\begin{equation} \label{eq:D_def}
D_{W|\bar{W}} (R) \triangleq \limsup_{T \rightarrow \infty} D_T(R),
\end{equation}
where 
\[
D_T(R) =  \inf \frac{1}{T} \int_0^T \mathbb E \left(W_t-\widehat{W}(t) \right)^2 dt,
\]
and the infimum is taken over all joint distributions $ \mathrm P_{\bar{W}^{N_T}, \widehat{W}^T}$ such that their mutual information rate \cite{pinsker1964information}
\begin{equation}
\label{eq:mutual_information}
T^{-1} I \left(\bar{W}^{N_T} ;  \widehat{W}^T \right)
\end{equation}
is limited to $R$ bits per unit time, and their marginal $P_{\bar{W}^{N_T}}$ coincides with the distribution of the samples of $W_{(\cdot)}$ over $[0,T]$. Then 
\[
D(f_s,R) = D_{W|\bar{W}}(R). 
\]
\end{thm}

\begin{IEEEproof}
For $T>0$, define the following distortion measure on $\mathbb R^{N_T} \times \Ltwo [0,T]$:
\begin{align}
\bar{d} & \left( \bar{w}^{N_T}, \widehat{w}^T \right)  \triangleq  \mathbb E \left[\frac{1}{T}\int_0^T (W_t-\widehat{w}_t)^2 dt | \,  \bar{W}^{N_T} = \bar{w}^{N_T}  \right]. \label{eq:amended_distortion} 
\end{align}
Here $\Ltwo [0,T]$ is the space of square integrable functions over $[0,T]$, $\widehat{w}^{T}$ is an element of this space, and the relation between $W^T$ and $\bar{W}^{N_T} $ is the same as in \eqref{eq:sampling_relation}. In words, $\bar{d}$ is the averaged quadratic distortion between the reconstruction waveform $\widehat{w}^T$ and all possible realizations of the random waveform $W^T$ whose values at the points $0,T_s,\ldots, N_T  T_s$ are given by $\bar{w}$. By properties of conditional expectation we have
\[
\mathbb E \bar{d}\left(\bar{W}^{N_T},\widehat{W}^T  \right) = \frac{1}{T} \int_0^T \mathbb E \left( W_t - \widehat{W}_t \right)^2 dt. 
\]
From the source coding theorem for i.i.d.\ random variables over arbitrary alphabets with a single-letter distortion measure \cite{Berger1968254}, it follows that the OPTA in encoding $\bar{W}^{N_T}$ is obtained by minimizing over all joint probability distributions of $\bar{W}^{N_T} $ and $\widehat{W}^T$ such that the mutual information rate \eqref{eq:mutual_information} is limited to $R$ bits per unit time. In the context of our problem, this source coding theorem implies an information representation for the OPTA under sampling at rate $f_s$ of an information source consisting of multiple independent realizations of the waveform $W^T$. Since we are interested in describing a single realization of $W^T$ as $T \to \infty$, what is required is an argument that allows us to separate the path of $W_{(\cdot)}$ over an arbitrary finite time into multiple sections (blocks), and consider the joint encoding thereof as multiple realizations over a fixed-length finite interval. \par
When the continuous source is ergodic or, more generally, \emph{asymptotic mean stationary}, such an argument is achieved by mixing properties of the probability space \cite{gray2011entropy}. In our case, however, $W_{(\cdot)}$ is not asymptotic mean stationary since its variance diverges, 
so a different argument must be used in order to separate the waveform into multiple i.i.d.\ sections in order to encode $W_{(\cdot)}$ over blocks. Such an argument was proposed by Berger \cite{berger1970information}: use a separate bitstream to encode the endpoints of all length-$T$ intervals. This task is equivalent to encoding a discrete-time Wiener process of variance $T\sigma^2$. It follows from \cite[Eq. 39]{fine1968response} that the distortion $\delta$ in an encoding of the latter using a delta modulator with $\bar{R}=RT$ bits per sample is smaller than a constant times $\sigma^2/\bar{R}$. For any finite $R$, this number can be made arbitrarily small by taking the blocklength $T$ large enough. That is, the endpoints $\bar{W}_{N_T}, \bar{W}_{2N_T},\ldots,$ can be described with high accuracy using an arbitrarily small number of bits per unit time. Since the increments of $W_{(\cdot)}$ are independent, its statistics conditioned on the sequence of endpoints is the same as of multiple i.i.d.\ realizations of $W^T$, and Theorem~\ref{thm:coding_theorem} follows from the first part of the proof.
\end{IEEEproof}

The representation \eqref{eq:separation} implies that $D(f_s,R)$ can be found by evaluating $D_{\tilde{W}}(R)$. An information theoretic expression for the latter follows from Theorem~\ref{thm:coding_theorem}:
\begin{cor} \label{cor:source_coding2}
Fix $R>0$ and $f_s>0$. Then
\begin{equation}
\label{eq:source_coding2}
D_{\tilde{W}}(R) = \limsup_{T\rightarrow \infty} \inf_{P_{\tilde{W}^T,\widehat{W}^T}} \frac{1}{T} \int_0^T \mathbb E \left(\tilde{W}_t - \widehat{W}_t \right)^2,
\end{equation}
where the infimum is taken over all joint distributions $ \mathrm P_{\tilde{W}^T,\widehat{W}^T}$ with mutual information not exceeding $TR$ bits, and whose marginal $\mathrm  P_{\tilde{W}^T}$ coincides with the distribution of $\tilde{W}_{(\cdot)}$ of \eqref{eq:W_tilde_def}. 
\end{cor} 
\begin{IEEEproof}
From the properties of optimal MSE estimation, $D_T(R)$ of Theorem~\ref{thm:coding_theorem} can be written as 
\begin{equation} \label{eq:source_coding2_proof}
D_T(R) = \mmse(W^T|\bar{W}^{N_T}) + \inf \frac{1}{T} \int_0^T\mathbb E \left(\tilde{W}_t - \widehat{W}_t \right)^2dt,
\end{equation}
with optimization over joint distributions as specified in Theorem~\ref{thm:coding_theorem}. Since the mutual information is invariant to invertible transformations of the random vector $\bar{W}^{N_T}$, and since $\tilde{W}^T$ is obtained from $\bar{W}^{N_T}$ by such a transformation, the optimization in \eqref{eq:source_coding2_proof} can be replaced by an optimization over joint distributions $P_{\tilde{W}^T,\widehat{W}^T}$ with mutual information  rate not exceeding $TR$ bits, and whose marginal $P_{\tilde{W}^T}$ coincides with the distribution of $\tilde{W}_{(\cdot)}$. Corollary~\ref{cor:source_coding2} now follows from \eqref{eq:separation} and since $D(f_s,R)=\limsup_{T\rightarrow \infty} D_T(R)$ by Theorem~\ref{thm:coding_theorem}.
\end{IEEEproof}

Next, we derive $D_{\tilde{W}}(R)$ and $D(f_s,R)$ in closed form by solving the optimization problem in \eqref{eq:source_coding2} and evaluating its limit as $T \to \infty$.  

\def\f{5.4/2-(3-2.7)*(5.4/2-2)}
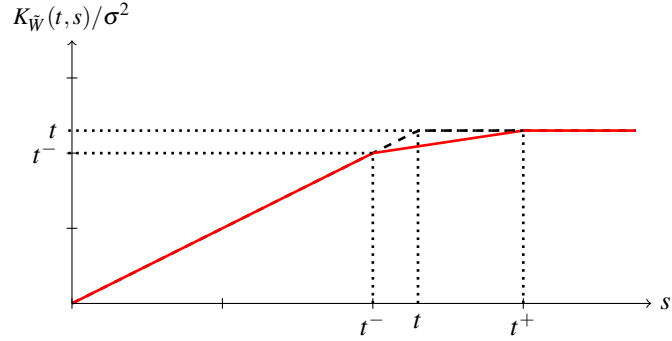
\begin{figure}
\begin{center}
\begin{tikzpicture}
\draw[dashed,line width = 1pt] plot[domain=0:7.5, samples=100] (\x,{min(\x/2,2.3)});
\draw[red, line width = 1pt]  plot[domain=0:4, samples=100] (\x,{\x/2});	
\draw[red, line width = 1pt]  plot[domain=4:6, samples=100] (\x,{\x/2-(3-2.3)*(\x/2-2)});
\draw[red, line width = 1pt]  plot[color=red,domain=6:7.5, samples=100] (\x,2.3);

\foreach \x in {0,2,4,6}
 \draw[shift={(\x,0)}] (0pt,2pt) -- (0pt,-2pt);
  \foreach \y/\ytext in {1/,2/,3/}
  \draw[shift={(0,\y)}] (2pt,0pt) -- (-2pt,0pt) node[left] {$\ytext$};
  \draw[->] (0,0) -- (0,3.5) node[above] {\small $K_{\tilde{W}}(t,s)/\sigma^2$};
    \draw[->] (0,0) --  (7.7,0) node[right] {$s$};
    \draw[dotted,line width = 1pt] (6,0) node[below] {$t^+$} -- (6,2.3);
    \draw[dotted,line width = 1pt] (4,0) node[below] {$t^-$} -- (4,2);
    \draw[dotted, line width = 1pt] (4.6,0) node[below] {$t$} -- (4.6,2.3);    
    \draw[dotted, line width = 1pt] (-0.05,2) node[left] {$t^-$}  --(4,2);
        \draw[dotted, line width = 1pt] (-0.05,2.3) node[left] {$t$} --(6,2.3);

\end{tikzpicture}
\caption{\label{fig:cov_non_diff} The autocovariance function of the process $\tilde{W}_{(\cdot)}$ for a fixed $t \in (0,T)$. }
\end{center}
\end{figure}

\subsection{The DRF of $\tilde{W}_{(\cdot)}$}
We use the KL expansion of $\tilde{W}_{(\cdot)}$ to evaluate $D_{\tilde{W}}(R)$. We have 
\[
W_t = \tilde{W}_t + B_t,\quad t\geq 0,
\]
where $B_{(\cdot)}$ and $W_{(\cdot)}$ are independent processes. The covariance function of $\tilde{W}_{(\cdot)}$ is given by 
\begin{equation} \label{eq:cov_W_tilde}
K_{\tilde{W}}(t,s) = K_W(t,s) - K_B(t,s). 
\end{equation}
The function $K_{\tilde{W}} (t,s)$ is illustrated for a fixed $t\in(0,T)$ in Fig.~\ref{fig:cov_non_diff}. Corollary~\ref{cor:source_coding2} implies that $D_{\tilde{W}}(R)$ is given as the limit in $T$ of the second term in \eqref{eq:source_coding2_proof}. Becasue $\tilde{W}_{(\cdot)}$ is a Gaussian process, this term is obtained by waterfilling over the eigenvalues in its KL transform \cite{berger1971rate}. These KL eigenvalues $\{\lambda_k,\,k=1,2,\ldots\}$ and their corresponding eigenfunctions $\{\phi_k, \,k=1,2,\ldots\}$ satisfy the Fredholm integral equation of the second kind \cite{zemyan2012classical}:
\begin{align} \label{eq:KL_integral}
\lambda_k \phi_k(t) & = \int_0^T K_{\tilde{W}} (t,s) \phi_k(s) ds.
\end{align}

Since $\tilde{W}^T$ is a linear combination of at most $N_T$ elements, its kernel defined by its autocovariance function is of rank at most $N_T$. We show in Appendix~\ref{app:eigenvalues} that $N_T$ of the eigenvalues of $K_{\tilde{W}} (t,s)$ satisfying \eqref{eq:KL_integral} are given by
\begin{equation}
\label{eq:eigenvalues}
\lambda_k = \frac{\sigma^2 T_s^2}{6}  \frac{ \left(2 \cos(k \pi )-\sin\left(\frac{(2 k-1) (N-1) \pi }{2N}\right)\right)}{ \left(\cos(k \pi )+\sin\left(\frac{(2 k-1) (N-1) \pi }{2 N}\right)\right)},\quad k=1,\ldots,N_T,
\end{equation}
and thus are the only eigenvalues of \eqref{eq:KL_integral}. We also show in Appendix~\ref{app:eigenvalues} that as $T$ goes to infinity with the ratio $f \triangleq k/T \approx k f_s/N_T$ kept constant for $0<f<f_s$, the density of these eigenvalues converges to the function
\begin{equation} \label{eq:density}
T_s^2\left(S_{\bar{W}}\left( \frac{\pi f}{2 f_s} \right)-\frac{1}{6} \right),\quad 0 < f < f_s,
\end{equation}
where $S_{\bar{W}}(\phi)$ is given in \eqref{eq:S_bar}. Existence of this limiting eigenvalue density implies the following result:
\begin{thm}  \label{thm:drf_w_tilde}
The DRF of the process $\tilde{W}_{(\cdot)}$, obtained by linearly interpolating the samples of a Wiener process at sampling rate $f_s$, is given by the following parametric expression:
\begin{subequations}
\label{eq:drf_wtilde}
\begin{align}
D_{\tilde{W}}(R_\theta) & =  \frac{\sigma^2}{f_s} \int_0^{1} \min \left\{\theta, S_{\bar{W}}(\phi) - \frac{1}{6} \right\} d \phi, \label{eq:drf_wtilde_D} \\
R_{\theta} & = \frac{f_s}{2} \int_{0}^{1} \log^+ \left[ \left( S_{\bar{W}}(\phi) - \frac{1}{6} \right)/\theta \right] d\phi \label{eq:drf_wtilde_R}, 
\end{align}
\end{subequations}
where $S_{\bar{W}}(\phi)$ is the limiting density of the eigenvalues in the KL expansion of $\bar{W}_{[\cdot]}$ given by \eqref{eq:S_bar}. 
\end{thm}
\begin{IEEEproof}
The density function \eqref{eq:density} satisfies the conditions of \cite[Thm. 4.5.4]{berger1971rate} (note that the stationarity property of the source is only needed in \cite[Thm. 4.5.4]{berger1971rate} to establish the existence of a density function, which in our case is given explicitly by \eqref{eq:density}). This theorem implies that the waterfilling expression over the eigenvalues $\{ \lambda_k\}_{k=1}^{N_T}$ converges, as $T$ goes to infinity, to the waterfilling expression over the density $S_{\tilde{W}}(f)$.
\end{IEEEproof}

We remark that as $f_s$ goes to infinity, $D_{\tilde{W}}(R)$ converges to $D_W(R)$ as can be seen by eliminating $\theta$ from \eqref{eq:drf_wtilde} in a similar way as in \eqref{eq:dr_asym}. In fact, this convergence already follows from the information representation of $D_{\tilde{W}}(R)$ and $D_W(R)$ in Corollary~\ref{cor:source_coding2} and \cite[Sec. IV]{berger1970information}, respectively,  
even without obtaining $D_{\tilde{W}}(R)$ in a closed form. Indeed, the kernel $K_{\tilde{W}}(t,s)$ converges to the kernel $K_{W}(t,s)$ in the $\Ltwo[0,T] \times \Ltwo[0,T]$ sense. As a result, the corresponding sequence of compact integral operators defined by \eqref{eq:KL_integral} converges, in the strong operator norm, to the operator defined by $K_{W}(t,s)$, showing that the eigenvalues \eqref{eq:eigenvalues} converge to the eigenvalues of the KL expansion for the Wiener process uniformly in $T$. This convergence of the eigenvalues implies convergence of $D_{\tilde{W}}(R)$ to $D_W(R)$, since both can be defined in terms of a uniformly bounded function of the eigenvalues of $K_{\tilde{W}}(t,s)$ and $K_{{W}}(t,s)$, respectively. Similar results were derived for cyclostationary Gaussian stationary processes in \cite{KipnisCyclo}.\\

From a practical point of view, it is important to emphasize that although $\tilde{W}_{(\cdot)}$ is a continuous-time process, its KL coefficients can be obtained directly from the discrete-time samples $\bar{W}_{[\cdot]}$ and without performing any analog integration as opposed to the KL coefficients of $W_{(\cdot)}$ in \eqref{eq:KL_Wiener}. Indeed, assuming for simplicity that $Tf_s$ is an integer, any integrable function $g(t)$ satisfies
\begin{align} \label{eq:KL_linear}
\int_0^T g(u) \tilde{W}_{u} du = \sum_{n=0}^{ f_sT-1} \left\{ \bar{W}_nX_n + \bar{W}_{n+1}Y_n \right\},
\end{align}
where
\begin{align*}
X_n & = \frac{1}{T_s} \int_{nT_s}^{(n+1)T_s } g(u) \left((n+1)T_s-u \right) du,\\
Y_n & = \frac{1}{T_s} \int_{nT_s}^{(n+1)T_s } g(u) \left(u-n T_s \right) du, \\
\end{align*}
By taking $g(t)$ to be the $k$th eigenfunction in the KL decomposition of $\tilde{W}_{(\cdot)}$ as given in Appendix.~\ref{app:eigenvalues}, we see that the $k$th KL coefficient of $\tilde{W}_{(\cdot)}$ over $[0,T]$ can be expressed as a linear function of the samples $\bar{W}^{N_T}$. This last fact implies that, in contrast to the achivable scheme in \cite{berger1970information}, a source code which is based on the KL transform of $\tilde{W}_{(\cdot)}$ may be applied directly to a linear transformation of $\bar{W}^{N_T}$ and does not require analog integration as in \eqref{eq:KL_Wiener}. \\

\subsection{The DRF of the Wiener Process given its Samples}
We are now ready to derive a closed-form expression for $D(f_s,R)$.
\begin{thm} \label{thm:main}
The indirect DRF of the Wiener process $W_{(\cdot)}$ given its uniform samples at rate $f_s$ and bitrate $R$ is given by the following parametric form:
\begin{subequations}
\begin{align}
D(R_\theta) & =  \frac{\sigma^2}{6f_s} + \frac{\sigma^2}{f_s} \int_0^1 \min \left\{\theta, S_{\bar{W}}(\phi) - \frac{1}{6}  \right\} d \phi, \label{eq:main_D} \\
R_\theta &= \frac{f_s}{2} \int_0^1 \log^+ \left[ \left(S_{\bar{W}}(\phi) - \frac{1}{6}\right) /\theta \right] d\phi \label{eq:main_R}.
\end{align} 
\label{eq:main}
\end{subequations}
\end{thm} 

\begin{IEEEproof}
Expression \eqref{eq:main} follows directly from \eqref{eq:mmse_value}, \eqref{eq:separation}, and Theorem \ref{thm:drf_w_tilde}.
\end{IEEEproof}

\begin{figure}
\begin{center}
\begin{tikzpicture}[xscale=1.3,yscale = 1.5]
\draw[<->,line width = 1pt, color = red] 
(0.290038/2, 1.85486)--(0.409617/2, 1.26863)--(0.527865/2, 0.951119) -- (0.679478/2, 0.706822) -- (0.90649/2, 0.495546) -- (1.26702/2, 0.318514) -- (1.26702/2, 0.318514) -- (1.0576, 0.147444) -- (1.27038, 0.107265) -- plot[domain=1.45:2.4, samples=15]  (\x, {  (2+sqrt(3))/6 * exp(-2*ln(2)*\x) });
\draw [dashed] (0.98,-0.1) node[below] {\small $\bar{R}_0$} -- (0.98,0.1667);
\draw[dashed] (-0.05,0.1666) node[left] {\small $\frac{1}{6}$} -- (0.98,0.1667);
\draw[->,line width=1pt] (0,0) -- (2.4,0) node[right, xshift=-0.1cm] {\small $\bar{R}$} ;
\draw[->,line width=1pt] (0,0) -- (0,2.5) node[above,xshift = -0.2cm] {\small $\widetilde{D}(\bar{R})$};   
\end{tikzpicture}
\begin{tikzpicture}[xscale=3,yscale = 2]
\def\th{0.72}
\draw[fill=blue!50] (0,0)--(0,\th)--(0.355,\th) -- plot[domain=0.355:1, samples=17]  (\x, {(1/(4*sin(deg(\x*1.571))*sin(deg(\x*1.571)) )-1/6 }) -- (1,0);

\draw[<-|,line width = 1pt] plot[domain=0.21:1, samples=15]  (\x, {   (1/(4*sin(deg(\x*1.571))*sin(deg(\x*1.571)) )-1/6 });

\draw[-|,dotted,line width = 1pt] plot[domain=0.23:1, samples=15]  (\x, { 1/( 4*sin(deg(\x*1.571))* sin(deg(\x*1.571 )) ) } );

\draw [dashed,line width=1pt]  (-0.05,\th) node[left] {\small $\theta$} --  (1,\th) ;
\draw [dashed,line width=0.5pt]  (-0.05,{ 1/( 4*sin(deg(1*3.14149/2))* sin(deg(1.571))) -0.1667 }) node[left] {\scriptsize $\frac{1}{12}$} --  (1,{1/12});
\draw [line width=1pt] (1,-0.1) node[below] {\scriptsize $1$} -- (1,0);
\draw [line width=1pt] (0,-0.1) node[below] {\scriptsize $0$} -- (0,0);
\draw[->,line width=1pt] (0,0) -- (1.05,0) node[right, xshift=-0.1cm] {\small $\phi$} ;
  \draw[->,line width=1pt] (0,0) -- (0,2);   
\node[above,xshift = -0.5cm,yshift = 0.2cm]  at (1,1/6)  {\scriptsize $S_{\bar{W}}(\phi)$};
\node[above,xshift = -0.2cm] at (0.3,2.2)  {\small $S_{\bar{W}}(\phi)-1/6$};
\end{tikzpicture}

\caption{\label{fig:waterfilling} Left: the function $\widetilde{D}(\bar{R})$ of \eqref{eq:D_tilde}. Right: waterfilling interpretation of the parametric equation \eqref{eq:D_tilde} describing $\widetilde{D}(\bar{R})$. }
\end{center}
\end{figure}
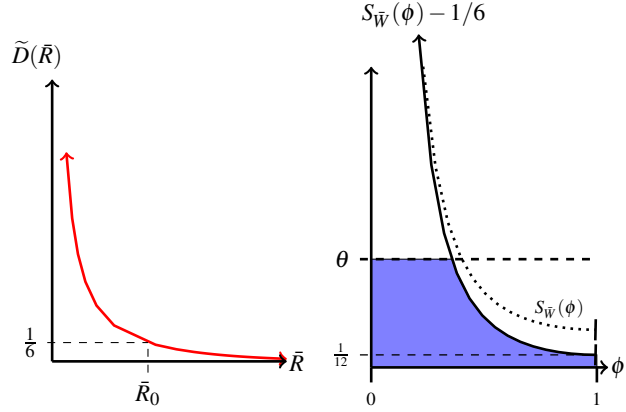

An alternative representation to \eqref{eq:main_D} is
\begin{align} \label{eq:DR_normalized}
D(f_s,R) = \frac{\sigma^2}{6f_s} + \frac{\sigma^2}{f_s} \widetilde{D} (\bar{R}),
\end{align}
where $\bar{R} = R/f_s$ and $\widetilde{D}(\bar{R})$ is given by
\begin{subequations}
\label{eq:D_tilde}
\begin{align} 
\widetilde{D}(R_\theta) & = \int_0^1\min \left\{\theta, S_{\bar{W}}(\phi)-\frac{1}{6} \right\} d\phi, \\
\bar{R}_\theta & = \frac{1}{2} \int_0^1 \log^+\left[ \left(S_{\bar{W}}(\phi)-\frac{1}{6} \right) / \theta \right] d\phi.
\end{align}
\end{subequations}
The function $\widetilde{D}(\bar{R})$ is dimensionless and only depends on the number of bits per sample $\bar{R}$. Figure~\ref{fig:waterfilling} illustrates $\widetilde{D}(\bar{R})$ and a  waterfilling interpretation of \eqref{eq:D_tilde}. \\ 

\begin{figure*}[tb!]
\begin{center}
\begin{tikzpicture}
\begin{loglogaxis}[
axis background/.style={fill=white},
ymajorgrids,
width=9cm, height=7cm,
xmin = 0.25, xmax=5, ymin = 0.05, ymax=1.15, 
samples=10, 
ylabel=MSE, 
xlabel= {$R$ [bits/time]},
xtick={0.5,1,1.5,2,2.5,3,3.5,4,4.5},
xticklabels={$0.5$,$1$,,$2$,,,,$4$,},
ytick={0.157,1},
yticklabels={$\frac{\sigma^2}{6f_s}$,$\sigma^2$},
line width=1.0pt,
mark size=1.5pt,
legend style= {at={(0.05,0.05)},anchor=south west,draw=black,fill=white,align=left},
]

\addplot[color = red, mark = x, solid, smooth, mark size=1.5pt] table [x=RRtrue, y=DDtrue, col sep=comma] {./PlotData/Wiener_DR_vs_R.csv};
\addlegendentry{$D(f_s,R)$};

\addplot[color = blue,no marks, smooth] table [x=RRdisc, y=DDce, col sep=comma] {./PlotData/Wiener_DR_vs_R.csv};
\addlegendentry{$D_{\CE}(R)$};

\addplot[color = black!30!green, mark = square*, mark size=1.5pt, solid, smooth, dashed, domain=0.25:5] {0.2924/x };
\addlegendentry{$D_W(R)$};

\addplot[color = black, no marks, dashed, smooth, domain=0.25:5, samples = 7] {0.157};
\addlegendentry{$\mmse(f_s)$};

\addplot[color =magenta, solid, smooth, dotted] table [x=RRdisc, y=DDdisc, col sep=comma] {./PlotData/Wiener_DR_vs_R.csv};
\addlegendentry{$D_{\bar{W}}(R/f_s)$};

\addplot[color = red, mark = x, solid, smooth] table [x=RRtrue, y=DDtrue, col sep=comma] {./PlotData/Wiener_DR_vs_R.csv};

\addplot[color = blue,no marks, smooth] table [x=RRdisc, y=DDce, col sep=comma] {./PlotData/Wiener_DR_vs_R.csv};

\coordinate (insetPosition) at (axis cs:0.75,0.5);
\end{loglogaxis}
\begin{loglogaxis} [at={(insetPosition)},
footnotesize,
width=5cm, height=2.5cm,
xmin = 0.75, xmax=3, 
ymin = 0.0005, ymax=0.02, 
ytick = {0.001, 0.01},
yticklabels = {\scriptsize $10^{-3}$,\scriptsize $10^{-2}$},
yticklabel pos=right,
xtick = {1,2},
xticklabels={\scriptsize $1$, \scriptsize $2$}, 
axis background/.style={fill=blue!5}]
]
\addplot[color = black, solid, smooth, mark size=1.5pt] table [x=R, y=Diff, col sep=comma] {./PlotData/Wiener_DR_vs_R_diff.csv};
\end{loglogaxis}
\end{tikzpicture}
\begin{tikzpicture}
\begin{loglogaxis}[
axis background/.style={fill=white},
ymajorgrids,
width=9cm, height=7cm,
xmin = 0.25, xmax=10, ymin = 0.05, ymax=1.15, 
samples=10, 
xlabel= {$f_s$ [1/time]},
xtick={0.25,0.5,1,1.5,2,2.5,3,3.5,4,4.5,5,5.5},
xticklabels={0.25,0.5,1,,2,,,,4,,,},
ytick={0.294,1},
yticklabels={$\frac{2\sigma^2}{\pi^2 \ln 2}$,$\sigma^2$},
line width=1.0pt,
mark size=1.5pt,
legend style= {at={(0.95,0.05)},anchor=south east,draw=black,fill=white,align=left},
]

\addplot[color = red, mark = x, solid, smooth, mark size=1.5pt] table [x=fs, y=DDtrue, col sep=comma] {./PlotData/Wiener_DR_vs_fs.csv};
\addlegendentry{$D(f_s,R)$};

\addplot[color = blue,no marks, smooth] table [x=fs, y=DDce, col sep=comma] {./PlotData/Wiener_DR_vs_fs.csv};
\addlegendentry{$D_{\CE}(R)$};

\addplot[color = black!30!green, mark = square*, mark size=1.5pt, solid, smooth, dashed, domain=0.25:5, samples = 7] {0.2924 };
\addlegendentry{$D_W(R)$};

\addplot[color = black, no marks, dashed, smooth, domain=0.25:5, samples = 7] {1/(6*x)};
\addlegendentry{$\mmse(f_s)$};

\addplot[color = magenta, solid, smooth, dotted] table [x=fs, y=DDdisc, col sep=comma] {./PlotData/Wiener_DR_vs_fs.csv};
\addlegendentry{$D_{\bar{W}}(R/f_s)$};

\addplot[color = blue,no marks, smooth] table [x=fs, y=DDce, col sep=comma] {./PlotData/Wiener_DR_vs_fs.csv};

\addplot[color = red, mark = x, solid, smooth] table [x=fs, y=DDtrue, col sep=comma] {./PlotData/Wiener_DR_vs_fs.csv};

\coordinate (insetPosition) at (axis cs:0.3,0.65);
\end{loglogaxis}
\begin{loglogaxis} [at={(insetPosition)},
footnotesize,
width=4.82cm, height=2.5cm,
xmin = 0.3, xmax=1.5, 
ymin = -0.01, ymax=0.02, 
ytick = {0.001, 0.01},
yticklabels = {\scriptsize $10^{-3}$,\scriptsize $10^{-2}$},
yticklabel pos=right,
xtick = {0.5,1},
xticklabels={\scriptsize $0.5$,\scriptsize $1$}, 
axis background/.style={fill=blue!5}]
ymajorgrids,
]
\addplot[color = black, solid, smooth, mark size=1.5pt] table [x=fs, y=Diff, col sep=comma] {./PlotData/Wiener_DR_vs_fs_diff.csv};
\end{loglogaxis}

\end{tikzpicture}

\caption{
The indirect DRF $D(f_s,R)$ of the Wiener process given its uniform samples and the compress-and-estimate upper bound $D_{\CE}(f_s,R)$, both as a function of: (left) $R$, with $f_s = 1$ samples per unit time, and (right) $f_s$, with $R = 1$ bits per unit time. Also shown are the DRF of the Wiener process $D_{W}(R)$, the DRF of the discrete-time Wiener process $D_{\bar{W}}(R/f_s)$, and the minimal MSE in estimating the Wiener process from its samples $\mmse(f_s)$. In both figures all axes have logarithmic scales and the inset shows the difference $D_{CE}(R,f_s) - D(f_s,R)$. 
\label{fig:RD_loglog} 
}
\end{center}
\end{figure*}
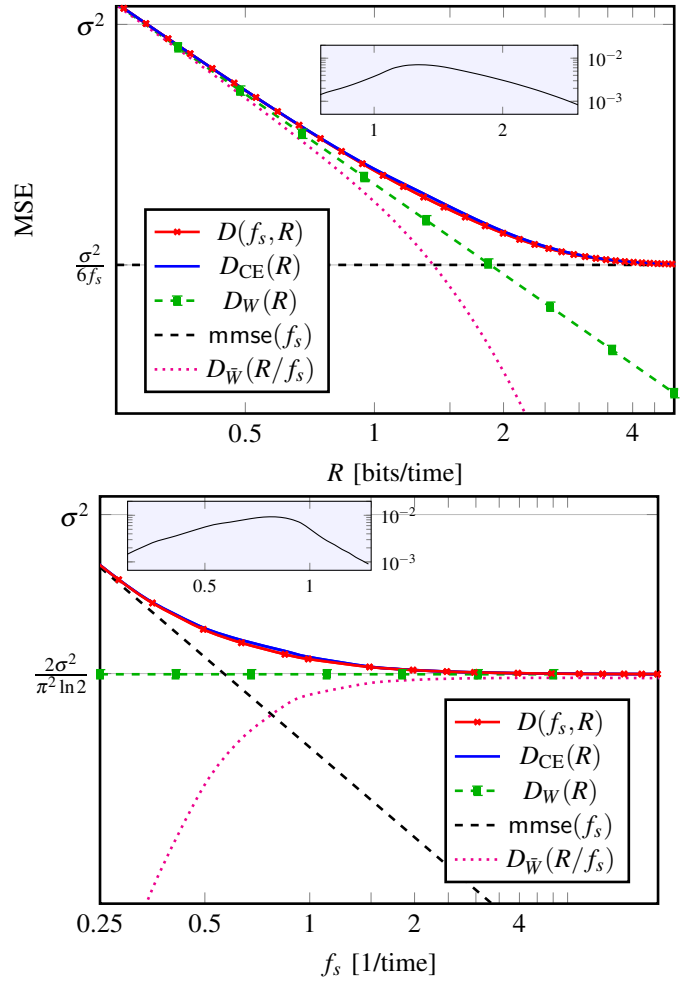

It follows from Theorem~\ref{thm:main} that $D(f_s,R)$ is monotonically decreasing in $f_s$ and converges to $D_{W}(R)$ as $f_s$ goes to infinity. We remark that monotonicity of $D(f_s,R)$ in $f_s$ is not evident in view of \cite[Exm. VI.2]{Kipnis2014}, where it is shown that the DRF of an arbitrary Gaussian stationary process given its samples may not be monotone in the sampling rate. Figure~\ref{fig:RD_loglog} illustrates expression \eqref{eq:main} along with other distortion functions defined in this paper, and confirms the anticipated behavior of $D(f_s,R)$ as $f_s$ or $R$ goes to infinity that is predicted in \eqref{eq:lim_fs} and \eqref{eq:R_to_infty}. 
We next study expression \eqref{eq:main} for $D(f_s,R)$ in the two regimes of low and high sampling rate $f_s$ compared to the bitrate $R$, corresponding to high and low bits per sample $\bar{R}$, respectively. 

\subsubsection{Low sampling rates}
As shown in Fig.~\ref{fig:waterfilling}, the minimal value of $S_{\bar{W}}(\phi)-1/6$, the integrand in \eqref{eq:main}, is $1/12$. Whenever
\begin{equation} \label{eq:border_point}
\frac{R}{f_s} \geq \frac{1+\log(\sqrt{3}+2)}{2} \approx 1.45,
\end{equation}
$\theta$ is smaller than $1/12$, in which case we can eliminate $\theta$ from \eqref{eq:main} and obtain
\begin{equation} \label{eq:idrf_low_rate}
D(f_s,R) = \frac{\sigma^2}{f_s} \left( \frac{1}{6}+ \frac{2+\sqrt{3}}{6} 2^{-2R/f_s} \right). 
\end{equation}

\subsubsection{High sampling rates}
When $R \ll f_s$, $\theta$ is large compared to $S_{\bar{W}}(\phi)-1/6$, and the integral in \eqref{eq:drf_wtilde_R} is non-zero only for small values of $\phi$. Using the Taylor expansion of $\sin^{-2}(x)$, we obtain
\begin{equation}
\label{eq:high_rate}
D(f_s,R) = \frac{2\sigma^2}{\pi^2 \ln 2} R^{-1} + \frac{\sigma^2 \ln 2}{18} \frac{R}{f_s^2} + O\left(f_s^{-4} \right). 
\end{equation}
From \eqref{eq:high_rate} we have that, as anticipated in \eqref{eq:lim_fs} and \eqref{eq:upper_bound}, $D(f_s,R)$ converges to $D_W(R)$ as $f_s \to \infty$. However, this rate of convergence is inversely quadratic in $f_s$, rather than the inverse linear convergence rate implied by the upper bound $D^U(f_s,R)$ from Corollary~\ref{cor:upper_bound}. 
\\

The behavior of $D(f_s,R)$ in the two cases above quantifies the intuitive fact that the distortion is dominated by the minimal MSE distortion $\mmse(f_s)$ 
for high values of bits per sample
$\bar{R}$, and by the lossy compression distortion  $D_{\tilde{W}}(R)$ for low values of $\bar{R}$. 
The transition between the two regimes occurs when the MMSE term in \eqref{eq:DR_normalized} equals the term $D_{\tilde{W}}(R)$ associated with lossy compression distortion, i.e., at some $\bar{R}_0$ satisfying $\widetilde{D}(\bar{R}_0) = 1/6$, which can be found to be $\bar{R}_0 \approx 0.98$. \\

The excess distortion in encoding the Wiener process at bitrate $R$ due to a rate $f_s$ sampling constraint is described by the ratio
\begin{equation} \label{eq:ratio_smp}
\rho_{\smp}(\bar{R}) \triangleq \frac{D(f_s,R)}{D_W(R)} =  \frac{\pi^2 \ln 2}{2} \left(\frac{1}{6}+ \widetilde{D}(\bar{R}) \right) \bar{R}. 
\end{equation}
Similarly, the excess distortion in sampling the Wiener process at rate $f_s$ due to a bitrate $R$ quantization or lossy compression constraint on the samples is described by the ratio
\begin{equation} \label{eq:ratio_qnt}
\rho_{\qnt}(\bar{R}) \triangleq \frac{D(f_s,R)}{\mmse(f_s)} =  1+ 6\widetilde{D}(\bar{R}). 
\end{equation}
Both $\rho_{\smp}(\bar{R})$ and $\rho_{\qnt}(\bar{R})$ are only a function of $\bar{R}$, implying that the performance loss due either to sampling or lossy compression are fully characterized by the average number of bits per sample consumed by the digital representation. As an example, given any source code for the samples of the Wiener process allocating $\bar{R}=1$ bits per sample on average, the distortion in recovering the process is at least
\[
\rho_{\smp}(1) D_W(R) \approx  1.18 D_W(R), 
\]
or 
\[
\rho_{\qnt}(1) \mmse(f_s) \approx 2.07 \mmse(f_s).  
\]
These numbers reflect $18\%$ loss compared to the optimal encoding without a sampling constraint, and $107\%$ loss compared to recovering the process from its samples without  quantizing them. \\

%

In the next section we study the distortion in recovering the Wiener process using a good sequence of encoders for the discrete-time process $\bar{W}_{[\cdot]}$, rather than an optimal sequence designed to attain $D_{\tilde{W}} (R)$. 

\section{Compress-and-Estimate \label{sec:ce} }

In Subsection~\ref{subsec:mmse} we concluded that the OPTA in the combined sampling and source coding problem is obtained via an EC source coding strategy: First estimate $W^T$ from the samples $\bar{W}^{N_T}$ resulting in $\tilde{W}^T$, and then compress $\tilde{W}^T$ using an optimal source code adjusted to its distribution. In this section we consider an alternative coding strategy: First encode the vector of samples $\bar{W}^{N_T}$ using an optimal source code, such that the expected MSE in recovering $\bar{W}^{N_T}$ from its encoded version converges to its DRF $D_{\bar{W}}(\bar{R})$ of \eqref{eq:DR_BM_disc}. Next, estimate $W^T$ from the the encoded representation of the samples using this code. We denote this scheme as compress-and-estimate (CE). See Fig.~\ref{fig:ec_vs_ce} for a block diagram of EC and CE.
\par
In this section we provide a precise characterization of the distortion under CE in the case where the encoding of the samples is done according to the scheme outlined in \cite{berger1970information} for attaining the DRF of the discrete-time Wiener process $\bar{W}_{[\cdot]}$. Specifically, we derive a distortion expression we denote as $D_{\CE}(f_s,R)$, and show this distortion is achievable using a sequence of codes whose bitrate converges to $R$ from above. We also show that when the bitrate is at most $R$, the distortion under this coding scheme is bounded from below by $D_{\CE}(f_s,R)$. Finally, by comparing $D_{\CE}(f_s,R)$ to $D(f_s,R)$, we conclude that the maximal ratio between the two is not greater than $1.027$, indicating a maximal performance penalty of $2.7\%$ in using CE over the optimal source coding scheme. 

\subsection{CE Encoding and Decoding \label{subsec:ce}}
Let $\{\bar{f}_N\}_{N \in \mathbb N}$ be a sequence of encoders indexed by their blocklength $N \in \mathbb N$. Assume that the encoder $\bar{f}_N$ operates according to the random coding scheme outlined in \cite{berger1970information} for achieving the DRF of the discrete-time Wiener process $\bar{W}_{[\cdot]}$. For completeness and further discussion, we now provide a detailed description of this scheme. \par
We describe the joint encoding of $L$ blocks of samples obtained over the time lag $TL$, so that each block contains roughly $N_T = \lfloor Tf_s \rfloor$ samples. Denote by $\Sigma_{\bar{W}}$ the covariance matrix of the vector $\bar{W}^{N_T}$ and consider the unitary matrix $\Um$ that satisfies 
\[
\Sigma_{\bar{W}} = \Um^T\Lambda \Um,
\] 
where $\Lambda$ is diagonal with the eigenvalue of $\Sigma_{\bar{W}}$ on its diagonal. These eigenvalues are given by \cite[Eq. 2]{berger1970information}
\[
\lambda_n = \frac{\sigma^2/f_s}{ 4\sin^2 \left( \frac{2n-1}{2N+1} \frac{\pi}{2} \right) },\quad n=1,\ldots,N_T,
\]
where their respective eigenvectors $u_1,\ldots,u_{N_T}$ are the columns of $\Um$. Given the $N_TL$ samples of $W^{TL}$, we consider the $N_T$ length $L$ sequences $B^{(1)},\ldots,B^{(N_T)}$, defined by
\begin{equation} \label{eq:KL_discrete}
B_l^{(n)} = u_n^T \bar{W}^{(l)} = \sum_{k=1}^{N_T} u_{k,n} \bar{W}^{(l)}_{k},\quad l=1,\ldots,L,\quad n=1,\ldots,N_T,
\end{equation}
where $\bar{W}^{(l)} \triangleq \bar{W}_{(l-1)N_T}^{lN_T} - \bar{W}_{(l-1)N_T}$. In words, $B_n^{(l)}$ is the $n$th coefficient in the KL decomposition of the $l$th $N_T$-length block of $\bar{W}^{L N_T}$, where this block is initialized so that $\bar{W}_0^{(l)}=0$. See Fig.~\ref{fig:CE_encoder} for an illustration of the relation between $B^{(n)}$ and $\bar{W}^{(l)}$. \par
Given a bitrate budget $R$, a bitrate slackness parameter $\rho>0$ and a blocklength $N_T$, we construct a codebook as follows: for each $n=1,\ldots,N_T$, we draw $2^{\lfloor{ (\bar{R}_n + \rho) L \rfloor} }$ codewords to describe $B^{(n)}$. Each codeword is obtained by $L$ independent draws from the scalar normal distribution with zero mean and variance $\left[\lambda_n - \theta \right]^+$. Here $\bar{R}_n$ and $\theta$ are determined by
\[
\bar{R}  = \frac{1}{N_T} \sum_{n=1}^{N_T} \bar{R}_n,
\]
where
\[
\bar{R}_n = \frac{1}{2} \log^+ \frac{\lambda_n}{\theta},
\]
and $\bar{R} \triangleq R / f_s$ is the number of bits per symbol. 
To each codeword $\widehat{b}^{(l)}$ we associate a unique index $i_n \in \left\{1,\ldots,2^{\lfloor R_n L \rfloor } \right\}$. We denote this codeword ensemble by $\mathcal C_n$, and reveal it to the decoder and decoder. Note that $\mathcal C_n$ is trivial whenever $\lambda_n  \leq  \theta$, since then $R_n = 0$. Therefore, in practice, we only need to consider the encoding of $B^{(1)}, \ldots, B^{(n_{\max} )}$ where $n_{\max}$ is the largest integer such that $\lambda_n > \theta$. \par
The encoding of a realization $\bar{w}^{LN_T}$ of $\bar{W}^{L N_T}$ is as follows: First divide $\bar{w}^{LN_T}$ into $L$ blocks of length $N_T$ each: $\bar{w}^{(1)},\ldots,\bar{w}^{(L)}$. Then obtain the $N_T$ length-$L$ sequences $b^{(1)},\ldots b^{(N_T)}$ from these blocks using
 \eqref{eq:KL_discrete}. For each $n=1,\ldots,N_T$, we associate the index $i_n$ corresponding to the codeword $\widehat{b}^{(n)}(i_n)$ of smallest Euclidean distance from $b^{(n)}$ in 
$\mathcal C_n$. The encoder outputs the indices $(i_1,\ldots,i_{N_T})$. In parallel to the representation of the block $\bar{W}^{LT}$ using $(i_1,\ldots,i_{N_T})$, in order to control the error due to uncertainty in  block starting locations, the encoder sends a separate 
bitstream obtained using a delta modulator applied to the sequence of block starting points. As explained in \cite[Sec. IV]{berger1970information}, the bitrate required for this representation goes to zero as $T$ goes to infinity, and hence the total rate of the code we described
is $R + o(1)$ where $o(1)$ goes to zero as both $L$ and $T/L$ go to infinity. 
\begin{figure}
\begin{center}
\begin{tikzpicture}
\draw(-0.75,3.75) -- node[above, rotate = -45, xshift = -0.2cm] { \small time} 
node[below, rotate = -45, xshift = -0.2cm] {\small block} 
(0,3);

\draw[fill = red!30] (0,0) rectangle (4,3);
\draw (0,0) rectangle (6,3);

\draw (1,0) --
 (1,3.5) node[above, xshift = -0.3cm, yshift = -0.5cm] { \small $B^{(1)}$}
 node[below, above, xshift = -0.5cm, yshift = -1.1cm] 
{\small $u_1^T W^{(1)}$} 
 node[below, above, xshift = -0.5cm, yshift = -1.6cm] 
{\small $u_1^T W^{(2)}$} 
 node[below, above, xshift = -0.5cm, yshift = -3.6cm] 
{\small $u_1^T W^{(L)}$} ;

\draw (2,0) -- (2,3.5) node[above, xshift = -0.3cm, yshift = -0.5cm] {\small  $B^{(2)}$} 
 node[below, above, xshift = -0.5cm, yshift = -1.1cm] 
{\small $u_2^T W^{(1)}$}
 node[below, above, xshift = -0.5cm, yshift = -1.6cm] 
{\small $u_2^T W^{(2)}$} 
 node[below, above, xshift = -0.5cm, yshift = -3.6cm] 
{\small $u_2^T W^{(L)}$} ;

\draw (4,0) -- (4,3.5) node[above, xshift = -0.5cm, yshift = -0.5cm] { \small $B^{(n_{\max} )}$} ;
\draw (3,0) -- (3,3.5);

\draw (4.9,0) -- (4.9,3.5) node[above, xshift = +0.5cm, yshift = -0.5cm] {\small  $B^{(N_T)}$} 
 node[below, above, xshift = +0.55cm, yshift = -1.1cm] 
{\small $u_{N_T}^T W^{(1)}$}
 node[below, above, xshift = +0.55cm, yshift = -1.6cm] 
{\small $u_{N_T}^T W^{(2)}$} 
 node[below, above, xshift = +0.55cm, yshift = -3.6cm] 
{\small $u_{N_T}^T W^{(L)}$} ;

\draw (0,2.5)  node[left, yshift = 0.25cm] {\small $l=1$}  --  (6,2.5);
\draw (0,2)  node[left, yshift = 0.25cm] {\small $l=2$}  -- (6,2);

\node at (-0.5,1.5) {$\vdots$};
\node at (0.5,1.5) {$\vdots$};
\node at (1.5,1.5) {$\vdots$};
\node at (2.5,0.25) {$\cdots$};

\node at (2.5,2.75) {$\cdots$};
\node at (2.5,2.25) {$\cdots$};

\node at (2.5,3.25) {$\cdots$};
\node at (4.5,3.25) {$\cdots$};
\draw  (0,0.5) node[left, yshift = -0.25cm] {\small $l=L$} -- (6,0.5);
\end{tikzpicture}
\caption{\label{fig:CE_encoder}
Description of encoding $\bar{W}^{LN_T}$ using the encoder $\bar{f}_{NL}$: divide the vector $\bar{W}^{LN}$ into $L$ blocks $\bar{W}^{(l)} = \bar{W}_{(l-1)N}^{lN}- \bar{W}_{(l-1)N}$, $l=1,\ldots,L$. For  $n=1,\ldots,N$, form the vector $B^{(n)}$ consisting of the $n$th coefficient in the KL transform of each of the $L$ blocks. Each such vector is encoded using a random Gaussian codebook of rate $L \bar{R}_n$ bits, where $\bar{R}_n = 0$ for $N>n_{\max}$. 
}
\end{center}
\end{figure}
%
%


We note that encoding using $\{\bar{f}_{N}\}_{N \in \mathbb N}$ corresponds to the achievability scheme outlined in \cite[Sec. IV]{berger1970information} for attaining the DRF of the discrete-time Wiener process $\bar{W}_{[\cdot]}$. That is, for any $\delta>0$ and $\rho>0$, there exists $N$ large enough such that 
\begin{equation}
\label{eq:source_coding_disct}
\mmse\left( \bar{W}^N  | \bar{f}_{N} \left(\bar{W}^N  \right) \right) - \delta < D_{\bar{W}}(\bar{R}). 
\end{equation}
In our case, however, we are interesting in recovering $W_{(\cdot)}$ rather than $\bar{W}_{[\cdot]}$, and hence the decoding in CE also involves the estimation of $W^T$ given the sequence consisting of decoded codewords. We now analyze the distortion with respect to $W_{(\cdot)}$ attained by using the sequence of encoders $\{ \bar{f}_N \}_{N \in \mathbb N}$ defined above. 

\subsection{Distortion Analysis}
In order to characterize the distortion by the coding scheme defined above, we define 
\begin{subequations}
\label{eq:Dce_def}
\begin{align} 
D_{\CE}(f_s,R_\theta) & = \frac{\sigma^2}{6f_s} + \frac{\sigma^2}{f_s} \int_0^1 \min\left\{\theta, S_{\bar{W}}(\phi)   \right\} \frac{S_{\bar{W}}(\phi) - \frac{1}{6} }{S_{\bar{W}}(\phi)} d\phi \label{eq:Dce_def_D}\\
R_\theta & =  \frac{f_s}{2} \int_0^1 \log^+ \left[ S_{\bar{W}} (\phi)/\theta\right] d\phi.
\end{align}
\end{subequations}
\begin{thm} \label{thm:ce_exact}
Fix $f_s$ and $R$. 
\begin{itemize}
\item[(i)]
There exist sequences $\{R_n \}_{n\in \mathbb N}$ and $\{ T_n \}_{n\in \mathbb N}$ with $R_n\rightarrow R$ and $T_n\rightarrow \infty$, such that, assuming 
$\bar{f}_{N_T}$ operates at rate $R_n$, we have
\[
\lim_{n \rightarrow \infty} \mmse\left(W^{T_n} | \bar{f}_{\lfloor T_n f_s\rfloor} \left(\bar{W}^{ \lfloor T_n f_s \rfloor } \right) \right) = D_{\CE}(f_s,R).
\]
\item[(ii)] 
For any $\epsilon>0$, there exists $T_0$ such that, for any $T>T_0$ and encoder $\bar{f}_{N_T}$, 
\[
\mmse\left(W^T | \bar{f}_{N_T}\left(\bar{W}^{N_T} \right) \right) \geq D_{\CE}(f_s,R).
\]
\end{itemize}
\end{thm}

\begin{IEEEproof}
See Appendix~\ref{app:proofs}. 
\end{IEEEproof}

Theorem~\ref{thm:ce_exact} says that when the samples of the Wiener process are encoded using a minimum distance encoder with respect to a random codebook drawn from the distortion-rate achieving distribution, the resulting distortion is asymptotically given by $D_{\CE}(f_s,R)$ of \eqref{eq:Dce_def}. 
In particular, since $\{\bar{f}_N \}_{N \in \mathbb N}$ defines a good sequence of codes with respect to $\bar{W}_{[\cdot]}$, Theorem~\ref{thm:ce_exact} tightens the upper bound of Corollary~\ref{cor:upper_bound}. Indeed, for any $R>0$ and $f_s>0$ we have
\[
D_{CE}(f_s,R) < D^U(f_s,R). 
\] 
 \par
 The expression $D_{\CE}(f_s,R)$ is illustrated in Fig.~\ref{fig:RD_loglog}. We now analyze it in the two regimes of high and low sampling rate compared to the bitrate, respectively: 
\subsubsection*{Low sampling rate} When $R \geq f_s$, \eqref{eq:Dce_def} reduces to  
\begin{equation}
\label{eq:two_thirds}
D_{\CE}(f_s,R) = \frac{\sigma^2}{6f_s}  + \frac{2}{3} D_{\bar{W}}(R/f_s) =  \frac{\sigma^2}{6f_s} + \frac{2}{3f_s} 2^{-2 R/f_s}. 
\end{equation}
Comparing \eqref{eq:two_thirds} with the optimal distortion in \eqref{eq:high_rate}, we have 
\begin{align*}
& D_{\CE}(f_s,R) - D(f_s,R)  = \frac{\sigma^2}{f_s} \frac{2-\sqrt{3}}{6} 2^{-2 R/f_s}, 
\end{align*}
whenever $R /fs \geq \left(1+\log(\sqrt{3}+2) \right)/2$.

\subsubsection*{High sampling rate}
Using the Taylor expansion of $\sin^{-2}(x)$, for $f_s \gg R$ we obtain
\begin{align*}
D_{\CE}(f_s,R) & = D_W(R) + \frac{7}{36} \frac{R \ln 2}{f_s^2} + O(f_s^{-4}),
\end{align*}
from which we conclude that, similarly to $D(f_s,R)$,  $D_{\CE}(f_s,R)$ converges to $D_W(R)$ in a rate inversely quadratic in $f_s$. \\

\begin{figure}
\begin{center}
\begin{tikzpicture}

\begin{loglogaxis}[
axis background/.style={fill=white},
ymajorgrids,
width=9cm, height=7cm,
xmin = 0.25, xmax=8, ymin = 0.99, ymax=1.04, 
samples=10, 
ylabel={\small $\frac{D_{\CE}(f_s,R)}{D(f_s,R)}$}, 
xlabel= {$\bar{R}$ [bit/smp]},
xtick={0.25,0.5,1,1.5,2,2.5,3,3.5,4,4.5,5,5.5},
xticklabels={0.25,0.5,1,,2,,,,4,,,},
ytick={1,1.027},
yticklabels={$1$, $1.027$},
line width=1.0pt,
mark size=1.5pt,
ymajorgrids,
legend style= {at={(0.95,0.05)},anchor=south east,draw=black,fill=white,align=left},
]

\addplot[color = blue, solid, smooth, mark size=1.5pt] table [x=Rbar, y=rho, col sep=comma] {./PlotData/Wiener_ratio.csv};

\end{loglogaxis}

\end{tikzpicture}

\caption{
The ratio $ D_{\CE}(f_s,R) / D(f_s,R)$ versus $\bar{R} = R/f_s$ describing the performance loss in using CE compared to the optimal source coding scheme. 
\label{fig:ratio} 
}
\end{center}
\end{figure}
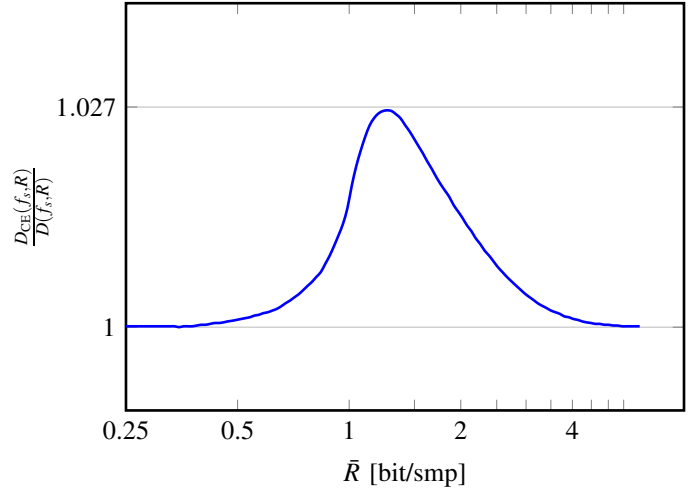

As in the case of $D(f_s,R)$, the excess distortion ratios $D_{\CE}(f_s,R)/D_W(R)$ and $D_{\CE}(f_s,R)/\mmse(f_s)$ are both only functions of the number of bits per sample $\bar{R}=R/f_s$. Therefore, the ratio between $D_{\CE}(f_s,R)$ and $D(f_s,R)$, describing the performance loss in using CE compared to the optimal scheme, also depends only in $\bar{R}$. As illustrated in Fig.~\ref{fig:ratio}, this ratio is bounded from above by $1.027$, indicating a maximal performance loss of only $2.7\%$ in using CE compared to the optimal source coding scheme.

%

\subsection{Understanding the Sub-optimality of CE}
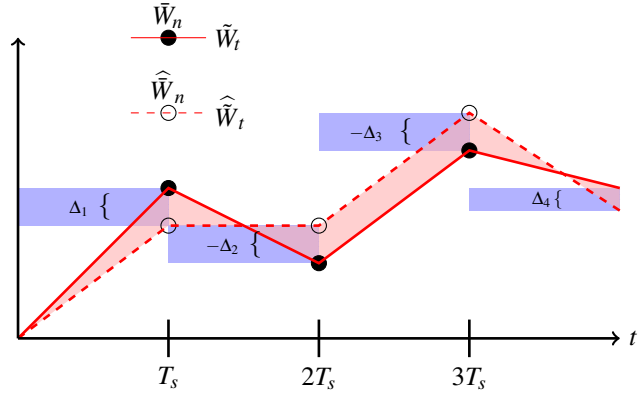
\begin{figure}
\begin{center}
\begin{tikzpicture}
\node[coordinate] (w1) at (2,2) {};
\node[coordinate] (w2) at (4,1) {};
\node[coordinate] (w3) at (6,2.5) {};
\node[coordinate] (w4) at (8,2) {};
\node[coordinate] (wh1) at (2,1.5) {};
\node[coordinate] (wh2) at (4,1.5) {};
\node[coordinate] (wh3) at (6,3) {};
\node[coordinate] (wh4) at (8,1.7) {};
\draw[line width = 1pt] (2,-0.25) node[below] {$T_s$} -- (2,0.25);
\draw[line width = 1pt] (4,-0.25) node[below] {$2T_s$} -- (4,0.25);
\draw[line width = 1pt] (6,-0.25) node[below] {$3T_s$} -- (6,0.25);

\fill[fill=blue!30, line width = 0pt] (w1) -- (wh1) -- +(-2,0) -- +(-2,0.5) -- (w1);
\fill[fill=blue!30, line width = 0pt] (w2) -- (wh2) -- +(-2,0) -- +(-2,-0.5) -- (w2);
\fill[fill=blue!30, line width = 0pt] (w3) --  (wh3) -- +(-2,0) -- +(-2,-0.5) -- (w3);
\fill[fill=blue!30, line width = 0pt] (w4) -- (wh4) -- +(-2,0) --  +(-2,+0.3) -- (w4);

\fill[fill=red!30, fill opacity=0.6] (w1) -- (wh1) -- (0,0) -- (w1);
\fill[fill=red!30, fill opacity=0.6] (w1) -- (wh1) -- (wh2) -- (w2) -- (w1);
\fill[fill=red!30, fill opacity=0.6] (w2) -- (wh2) -- (wh3) -- (w3) -- (w2);
\fill[fill=red!30, fill opacity=0.6] (w3) -- (wh3) -- (wh4) -- (w4) -- (w3);

\draw[fill = black] (w1) circle (0.1cm);
\draw[fill = black] (w2) circle (0.1cm);
\draw[fill = black] (w3) circle (0.1cm);
\draw[line width = 1pt,  color = red] (0,0) -- (w1) -- (w2) -- (w3) -- (w4);

\draw (wh1) circle (0.1cm);
\draw (wh2) circle (0.1cm);
\draw (wh3) circle (0.1cm);
\draw[line width = 1pt, dashed, color = red] (0,0) -- (wh1) -- (wh2) -- (wh3) -- (wh4);

\draw[line width = 0pt, dotted] (wh1) -- node[left, xshift = -0.6cm] { {\scriptsize $\Delta_1$} $\left\{ \right.$} (w1);
\draw[line width = 0pt, dotted] (wh2) -- node[left, xshift = -0.6cm] { {\scriptsize $-\Delta_2$}  $\left\{ \right.$} (w2);
\draw[line width = 0pt, dotted] (wh3) -- node[left, xshift = -0.6cm] {  {\scriptsize $-\Delta_3$ } $\left\{ \right.$} (w3);
\draw[line width = 0pt, dotted] (wh4) -- node[left, xshift = -0.6cm] {\scriptsize $\Delta_4 \left\{ \right.$} (w4);

\node[coordinate] (W_tilde) at (2,4) {};
\draw[fill = black] (W_tilde) circle (0.1cm);
\draw[color = red] (W_tilde)+(-0.5,0)-- (W_tilde) node[above, color = black] {$\bar{W}_n$} -- +(0.5,0) node[right, color = black] {$\tilde{W}_t$};

\node[coordinate] (W_hat) at (2,3) {};
\draw (W_hat) circle (0.1cm);
\draw[dashed, color = red] (W_hat)+(-0.5,0)-- (W_hat) node[above, color = black] {$\widehat{\bar{W}}_n$} -- +(0.5,0) node[right, color = black] {$\widehat{\tilde{W}}_t$};

\draw[->, line width = 1pt] (0,0) -- (8,0) node[right] {$t$} ;
\draw[->, line width = 1pt] (0,0) -- (0,4);

\end{tikzpicture}
\caption{The error in estimating $\bar{W}_{[\cdot]}$ and  $\tilde{W}_{(\cdot)}$ corresponds to the blue and red areas, respectively. Negatively correlated $\Delta_n$ and $\Delta_{n+1}$ are preferred for minimizing the distortion with respect to $\tilde{W}_{(\cdot)}$, while the distortion with respect to $\bar{W}_{[\cdot]}$ is indifferent to this correlation.
\label{fig:CE_suboptimality}}
\end{center}
\end{figure}

In order to gain some insight into the difference between the performance of CE compared to the optimal source coding scheme, it is useful to focus on the term 
\begin{equation}
\label{eq:delta_sum}
\frac{1}{N_T}\sum_{n=0}^{N_T-1} \mathbb E \Delta_n \Delta_{n+1}
\end{equation}
in the upper and lower bounds of Lemma~\ref{lem:bounds}. 
For simplicity, consider the regime $R\geq f_s$ in which we have
\begin{equation} \label{eq:two_thirds_2}
D_{\CE}(f_s,R) = \mmse(f_s) + \frac{2}{3} D_{\bar{W}}(f_s/R). 
\end{equation}
By evaluating \eqref{eq:lemma_exact_upper} in the limit $T\to \infty$ under the CE encoders $\{\bar{f}_N\}_{N\in \mathbb N}$ and comparing it with \eqref{eq:two_thirds_2}, it follows that \eqref{eq:delta_sum} goes to zero under CE. We now argue that, as opposed to CE, under the optimal encoder the term \eqref{eq:delta_sum} is negative. For this purpose, we illustrate in Fig.~\ref{fig:CE_suboptimality} the error term each encoder strives to minimize: error with respect to $\tilde{W}^T$ (red) for the optimal encoder, and error with respect to $\bar{W}^{N_T}$ (blue) for the CE encoder. An examination of these terms reveals that the latter is indifferent to the sign of $\Delta_n$, whereas the distortion with respect to $\tilde{W}^T$ is smaller whenever $\Delta_n$ and $\Delta_{n+1}$ alternate their signs (compare the red areas in the intervals $[T_s,2T_s]$ and $[2T_s,3T_s]$, corresponding to such sign alternation and no sign alternation, respectively). Therefore, an EC codebook favors a sign alternation from $\Delta_n$ to $\Delta_{n+1}$, implying that  \eqref{eq:delta_sum} is negative under EC. 

\section{Conclusions \label{sec:conclusion}}
We considered the minimal distortion in estimating the path of a continuous-time Wiener process from a bitrate-limited version of its uniform samples, taken at a finite sampling rate. We derived a closed form expression for the minimal distortion in this setting, given in terms of the asymptotic distribution of the KL eigenvalues of the estimator of the Wiener process from its samples. This expression allows us to determine the excess distortion in encoding the Wiener process under a uniform sampling constraint compared to its DRF, or, alternatively, the excess distortion in sampling the Wiener process under a quantization constraint compared to the MMSE from infinite precision samples. \par
In addition to the optimal source coding performance, we also consider a CE coding approach in which the Wiener process is estimated from an encoded version of its samples, where this encoding is chosen to minimize the distortion with respect to the samples rather than the continuous-time process. We provided a closed form expression for the performance under this approach, and showed that the performance loss under this sub-optimal approach is smaller than $2.7\%$ compared to the optimal source coding technique. 

\appendices

\section{ \label{app:eigenvalues}}
In this appendix we prove that the eigenvalues of the KL integral \eqref{eq:KL_integral} are given by \eqref{eq:eigenvalues}.  \\

Equation \eqref{eq:KL_integral} can be written as
\begin{align}
\frac{\lambda}{\sigma^2} \phi(t) &= \int_0^{t^-} s\phi(s)ds \nonumber \\
& + \int_{t^-}^{t^+}  \left(\frac{t-t^-}{T_s}s+t^-\frac{t^+-t}{T_s}  \right)\phi(s)ds  \label{eq:diff0} \\
& \quad + t \int_{t^+}^T \phi_k(s) ds. \nonumber
\end{align}
Differentiating the last expression leads to
\begin{align}
\frac{\lambda}{\sigma^2} \phi'(t) & =  
\int_{t^-}^{t^+} \frac{s-t^-}{T_s}\phi(s) ds  + \int_{t^+}^T \phi_k(s) ds,
 \label{eq:diff1}
\end{align}
which implies
\begin{equation} \label{eq:diff2}
\frac{\lambda}{\sigma^2} \phi''(t) = 0.
\end{equation}
We conclude that the solution to \eqref{eq:diff0} is a piece-wise linear function on intervals of the form $[nTs,(n+1)T_s)$ for $n=0,\ldots,N$,
where $N =  T/T_s $ (since the DRF is obtained by evaluating the solution as $T$ goes to infinity, and since this limit exists, there is no loss in generality by assuming $T/T_s$ is an integer), namely
\[
\phi_k(t) = \frac{t^+ - t}{T_s} a_k(t^-) + \frac{t-t^-}{T_s}b_k(t^-),\quad k =1,2,\ldots. 
\]
Equations \eqref{eq:diff0} and \eqref{eq:diff1} impose the following condition on the coefficients $a_k(t^-)$ and  $b_k(t^-)$, for $t \in [0,T_s N]$:
\begin{align*}
& \frac{\lambda}{T_s\sigma^2} \left( b_k(t^-)- a_k(t^-) \right) 
\\
& \quad = \frac{1}{6T_s} \left( a_k(t^-)+b_k(t^-) \right) + \frac{1}{2} \int_{t^+}^T \left( a_k(s^-) + b_k(s^-) \right) ds.
\end{align*}
\par

By imposing the initial conditions in \eqref{eq:diff0} and \eqref{eq:diff1}, it follows that the eigenfunctions in the KL transform are of the form
\begin{align*}
\phi_k(t) = \sqrt{A_k} \left(\frac{t^+-t}{T_s}\sin \left( \frac{2k-1}{2T}\pi t^- \right) + \right. \\
\quad 
\left.
\frac{t-t^-}{T_s} \sin \left( \frac{2k-1}{2T}\pi t^+ \right) \right),\quad k=1,\ldots,N,
\end{align*}
where $A_k$ is a normalization constant. The corresponding eigenvalues can be found by evaluating \eqref{eq:diff1}, which leads to 
\[
\lambda_k = \frac{\sigma^2 T_s^2}{6}  \frac{ \left(2 \cos(k \pi )-\sin\left(\frac{(2 k-1) (N-1) \pi }{2N}\right)\right)}{ \left(\cos(k \pi )+\sin\left(\frac{(2 k-1) (N-1) \pi }{2 N}\right)\right)},\quad k=1,\ldots,N.
\]

\section{ \label{app:proofs}}
In this append we provide the proofs of Lemma~\ref{lem:bounds}, Corollary~\ref{cor:upper_bound}, and Theorem \ref{thm:ce_exact}.

\subsection{Proof of Lemma~\ref{lem:bounds}}
Fix $T$, $f_s$ and let $\bar{R} = R/f_s$, $N = \lfloor Tf_s \rfloor$ and $M = f(\bar{W}^N)$. We have
\begin{equation}
\mmse\left(W^T|M \right) = \mmse\left(W^T | \bar{W}^N \right) + \mmse\left(\tilde{W}^T| M \right),
\label{eq:proof_five_2}
\end{equation}
hence we only focus on the term $\mmse\left(\tilde{W}^T| M\right)$. Denote $\widehat{\bar{W}}_n = \mathbb E \left[ \bar{W}_n | M \right]$. Then
\begin{align}
\mathbb E \left[ \tilde{W}_t | \widehat{\bar{W}}^N \right] & = \frac{t^+-t}{T_s} \mathbb E\left[W_{t^-}|\widehat{\bar{W}}^N \right] + \frac{t-t^-}{T_s} \mathbb E\left[W_{t^+}|\widehat{\bar{W}}^N \right] \nonumber \\
& = \frac{t^+-t}{T_s} \widehat{\bar{W}}_{f_st^-} + \frac{t-t^-}{T_s} \widehat{\bar{W}}_{f_st^+}, \label{eq:proof_bound_note}
\end{align}
Consider
\begin{align}
& \mmse \left(\tilde{W}^{NT_s}| \widehat{\bar{W}}^N \right)  = \frac{1}{NT_s}\int_0^{NT_s} \mathbb E \left(\tilde{W}_{t} - \mathbb E\left[\tilde{W}_t | \widehat{\bar{W}}^N\right] \right)^2 dt \nonumber \\
& = \frac{1}{NT_s} \sum_{n=0}^{N-1} \int_{nT_s}^{(n+1)T_s} \mathbb E \left(\tilde{W}_t-  \mathbb E \left[\tilde{W}_t | \widehat{\bar{W}}^N \right] \right)^2 dt \nonumber  \\
& \overset{a}{=} \frac{1}{NT_s} \sum_{n=0}^{N-1} \int_{nT_s}^{(n+1)T_s} \mathbb E \left(\tilde{W}_t- \frac{t-nT_s}{T_s} \widehat{\bar{W}}_{n+1} \ldots \right. \\
& \quad \quad \quad \quad \quad \quad \left. - \frac{T_s(n+1)-t}{T_s} \widehat{\bar{W}}_{n}  \right)^2 dt \nonumber  \\
& \overset{b}{=}\frac{1}{NT_s^3} \sum_{n=0}^{N-1}  \int_{nT_s}^{(n+1)T_s} \mathbb E \left( (t-nT_s) \Delta_{n+1}+((n+1)T_s-t) \Delta_n \right)^2 dt,  \label{eq:proof_five_1}
\end{align}
where $(a)$ follows from \eqref{eq:proof_bound_note}, $(b)$ follows since 
\begin{align*}
\tilde{W}_t & = \frac{t-t^-}{T_s}W_{t^+}+\frac{t^+-t}{T_s}W_{t^-} \\
& =\frac{t-t^-}{T_s} \bar{W}_{f_st^+}+\frac{t^+-t}{T_s}\bar{W}_{f_st^-},
\end{align*}
and by introducing the notation
\[
\Delta_n \triangleq \bar{W}_n - \widehat{\bar{W}}_n = \bar{W}_n - \mathbb E \left[\bar{W}_n | M \right] .
\]
Evaluating the integral in \eqref{eq:proof_five_1} we obtain
\begin{align*}
& \mmse(\tilde{W}^{NT_s} | M)  = \frac{1}{N} \sum_{n=0}^{N-1}  \left( \frac{1}{3} \mathbb E \Delta_{n+1}^2 + \frac{1}{3} \mathbb E\Delta_n^2 + \frac{1}{3} \mathbb E\Delta_{n+1} \Delta_n \right) \nonumber \\
& =  \frac{2}{3} \frac{1}{N} \sum_{n=1}^{N-1}  \mathbb E \Delta_n^2  +  \frac{1}{3N}  \sum_{n=0}^{N-1} \mathbb E\Delta_n \Delta_{n+1} + \frac{1}{3} \mathbb E \Delta_N^2 \\
& \geq \frac{2}{3} \frac{1}{N} \sum_{n=1}^{N-1}  \mathbb E \Delta_n^2  +  \frac{1}{3N}  \sum_{n=0}^{N-1} \mathbb E\Delta_n \Delta_{n+1},
\end{align*}
where we used the fact that $\mathbb E \Delta_0 =0$ because $W_0=0$ with probability one. Similarly, we have
\begin{align*}
& \mmse(\tilde{W}^{(N+1)T_s} | M) \nonumber \\
&  = \frac{2}{3} \frac{1}{N+1} \sum_{n=0}^N  \mathbb E \Delta_n^2  +  \frac{1}{3(N+1)} \sum_{n=0}^N \mathbb E\Delta_n \Delta_{n+1} +  \frac{1}{3} \mathbb E\Delta_{N+1}^2 \\
& \leq \frac{2}{3} \frac{1}{N+1} \sum_{n=1}^{N+1}  \mathbb E \Delta_n^2  +  \frac{1}{3(N+1)} \sum_{n=1}^N \mathbb E\Delta_n \Delta_{n+1}.
\end{align*}
The bounds \eqref{eq:lemma_exact_lower} and \eqref{eq:lemma_exact_upper} follow from the last two inequalities and the fact that
\[
\mmse(\tilde{W}^{NT_s}|M) \leq \mmse(\tilde{W}^T|M) \leq \mmse(\tilde{W}^{(N+1)T_s}|M). 
\]

\subsection{Proof of Corollary~\ref{cor:upper_bound} }
Set $N = \lfloor T f_s \rfloor $ and $\bar{R} = R/f_s$. By bounding $\mathbb E \Delta_{n+1}\Delta_n$ in \eqref{eq:lemma_exact_upper} from above by $
(\mathbb E \Delta_n^2 + \mathbb E \Delta_{n+1}^2)/2$, we obtain
\begin{align*}
&  \frac{1}{ N+1} \sum_{n=1}^N \mathbb E \Delta_n \Delta_{n+1} =  \frac{1}{ N+1} \sum_{n=0}^N \mathbb E \Delta_n \Delta_{n+1} \\
& \leq  \frac{1}{N+1} \sum_{n=0}^N \left( \frac{\mathbb E\Delta_n^2}{2} + \frac{\mathbb E\Delta_{n+1}^2}{2} \right) \\
& = 
\frac{1}{N+1} \sum_{n=1}^N \mathbb E \Delta_n^2 + \frac{1}{2} \frac{1}{N+1} \mathbb E \Delta_{N+1}^2  \leq \frac{1}{N+1} \sum_{n=1}^{N+1} \mathbb E \Delta_n^2. 
\end{align*}
It follows from \eqref{eq:lemma_exact_upper} that
\begin{align*}
&  D(f_s,R) \leq \mmse \left(W^T| \bar{f} (\bar{W}^N) \right)  \leq \mmse(W^T | \bar{W}^N) \\
& + \frac{2}{3} \frac{1}{N+1} \sum_{n=1}^{N+1} \mathbb E \Delta_n^2 +  
\frac{1}{3} \frac{1}{N+1} \sum_{n=1}^{N+1} \mathbb E \Delta_n^2 \\
& = \frac{1}{N+1} \sum_{n=1}^{N+1} \mathbb E \Delta_n^2 = \mmse( \bar{W}^{N+1} | \bar{f} \left( \bar{W}^N \right). 
\end{align*}
Under the  sequence of encoders $\left\{ \bar{f}_N,\, N \in \mathbb N\right\}$ in the limit $T\rightarrow \infty$, 
we have that $\mmse \left( \bar{W}^N | \bar{f}_N \left( \bar{W}^N \right) \right)$, 
and therefore $\mmse( \bar{W}^{N+1} | \bar{f}_N \left( \bar{W}^N \right)$, converge to $D_{\bar{W}}(\bar{R})$. In this limit we also have that $\mmse(W^T|\bar{W}^N )$ converges to $1/(6f_s)$, so that
\[
D(f_s,R) \leq \frac{1}{6f_s} + D_{\bar{W}} ( \bar{R}). 
\]
 
\subsection{Proof of Theorem~\ref{thm:ce_exact} }
For $L \in \mathbb N$ and $T>0$ we consider the encoding of the vector of samples $\bar{W}^{LT}$ using the encoders $\{\bar{f}_N \}_{N \in \mathbb N}$ and the estimation of $W^{LT}$ from this encoding. 
Throughout the proof we make use of various simplification for the notation in the paper, as per the following list:
\begin{itemize}
\item The distortion $D$ is normalized by $\sigma^2 / f_s$ and, consequently, assume that the any length $N_T$ vector $\bar{W}^{(l)}$ and its reconstruction $\widehat{\bar{W}}^{(l)}$ are normalized by $\sqrt{\sigma^2 / f_s}$. 
\item $N \triangleq  N_T \triangleq \lfloor T f_s \rfloor$
\item $\bar{f} \triangleq \bar{f}_{N}$ where the blocklength $N$ is understood from the context.
\end{itemize}

We first consider properties of the joint distribution that attains the DRF of the vector $\bar{W}^N$. For a prescribed $\bar{R}$, let $\theta$ be such that
\[
\bar{R} = \frac{1}{2} \sum_{k=1}^N \log^+ \left[ \lambda_k/\theta\right]. 
\]
Consider the eigenvalue decomposition of the matrix $\Sigma_{\bar{W}}$:
\[
\Sigma_{\bar{W}}= \Um^T \Lambda \Um,
\]
where $\Um$ is unitary and $\Lambda$ is diagonal. The elements $\lambda_1,\ldots,\lambda_N$ on the diagonal of $\Lambda$ are given by 
 \cite[Eq. 2]{berger1970information}
\[
\lambda_k  = \frac{1}{4\sin^2\left(\frac{2k-1}{2N+1} \frac{\pi}{2} \right)},\quad k=1,2,\ldots,N.
\]
The columns of $\Um$ are the eigenvectors in the KL transform of $\bar{W}^N$ corresponding to the eigenvalues $\lambda_1,\ldots,\lambda_N$, which are given by \cite{berger1970information}
\[
u_{k,n} = A_k \sin \left( \frac{2k-1}{2N+1} \pi n \right),
\]
and where $A_k$ is a normalization coefficient satisfying
\[
A_k  = \frac{1}{N} \sum_{n=0}^{N-1} \sin^2 \left(\frac{2k-1}{2N+1}  \pi n \right) = 1,\quad k = 1,2,\ldots,N.
\]

Given an encoder 
\begin{equation}
\label{eq:ce_proof_g}
g : \{ 1, \ldots, 2^{L \bar{R}_1 } \} \times \cdots \times \{1, \ldots,2^{L \bar{R}_{N_T }} \} \to \mathbb R^{[0,LT]},
\end{equation}
we denote 
\[
\widehat{W}^{LT} = g\left( f\left(\bar{W}^{LN_T} \right) \right),
\]
and 
\[
\Delta_n \triangleq W_n - \widehat{\bar{W}}_n, \quad n=1,\ldots,LN,
\]
where $\widehat{\bar{W}}_n = \widehat{W}_{n/f_s}$. \\

In order to prove (i), it is enough to show that for any $\rho>0$. $\epsilon>0$ and $\delta>0$, there exists $T$ and $L$ large enough and a decoder $g$ such that $L/T < \epsilon$, and, if $\bar{W}^{LN}$ is encoded using $f_{LN_T}$, then 
\[
\frac{1}{T} \int_0^T \mathbb E \left( W_t - [g\left( f\left( \bar{W}^{LT}\right) \right)]_t  \right)^2 dt < D_{\CE}(f_s,R) + \delta.
\]
(the condition $L/T < \epsilon$ is required to guarantee that the bitrate consumed by the delta modulator is arbitrarily small). In order to prove (ii), we show that for any $L$, and a function $g$ of the form \eqref{eq:ce_proof_g}, there exists $T_0$ such that 
\[
\frac{1}{T} \int_0^T \mathbb E \left( W_t - g\left( f\left( \bar{W}^{LT}\right) \right)  \right)^2 dt \geq D_{\CE}(f_s,R),
\]
whenever $\rho=0$ and $T\geq T_0$.  \\

We first prove the following claims:
\begin{enumerate}
\item[I.] Under the sequence of encoders $\{\bar{f}\}$, 
\begin{align}
& \frac{1}{NL} \sum_{n=1}^{LN}  \mathbb E \Delta_n \Delta_{n+1}  \nonumber \\
& = \frac{1}{N} \sum_{n=1}^{N}  \sum_{k=1}^{N} u_{n,k} u_{n+1,k} \frac{1}{L} \sum_{l=1}^L \mathbb E  \left[  \left(B_l^{(k)} - \widehat{B}_l^{(k)} \right)^2 \right]. \label{eq:ce_proof_step_ii}
\end{align}
\item[II.] For any $N \in \mathbb N$ and $\bar{R}>0$, 
\begin{align}
\lim_{N\rightarrow \infty} \frac{1}{N}  \sum_{n=1}^{N-1}  \sum_{k=1}^N  u_{k,n} u_{k,n+1}
\min \left\{\lambda_k, \theta\right\}  =  D_{\bar{W}}(\bar{R}) - \frac{1}{2}G(\bar{R})
\label{eq:ce_proof_step_i}
\end{align}
where
\[
G(\bar{R}_\theta) = \int_0^1 \frac{\min \left\{ S_{\bar{W}}(\phi),\theta \right\} }{S_{\bar{W}}(\phi)}  d\phi. 
\]
\item[III.] For any $f_s$ and $R$, 
\[
D_{\CE}(f_s,R) = \frac{1}{6 f_s} + \frac{2}{3} D_{\bar{W}}(\bar{R})  + \frac{1}{3} \left( D_{\bar{W}}(\bar{R})- \frac{1}{2}G(\bar{R}) \right). 
\]
\end{enumerate}

\subsubsection*{Proof of Claim I}
For $l=1,\ldots,L$ denote $Y^{(l)} = \Um \bar{W}^{(l)}$ and $\widehat{Y}^{(l)} = \Um \widehat{\bar{W}}^{(l)}$. Note that $B^{(n)}_l = Y^{(l)}_n$ and, since $B^{(n)}$ and $\widehat{B}^{(n)}$ are independent from $B^{(k)}$ and $\widehat{B}^{(k)}$ for $k\neq n$, we have that 
\[
\mathbb E \left( Y^{(l)}_n - \widehat{Y}^{(l)}_n \right) \left( Y^{(l)}_k - \widehat{Y}^{(l)}_k \right) = 0,
\]
for $k \neq n$. Next,
\begin{align}
& \frac{1}{NL} \sum_{n=1}^{LN}  \mathbb E \Delta_n \Delta_{n+1} \nonumber \\
& = 
\frac{1}{LN} \sum_{n=1}^{N-1}  \sum_{l=1}^L \mathbb E \left( \bar{W}_{n}^{(l)} - \widehat{\bar{W}}_{n}^{(l)}  \right)  \left( \bar{W}_{n+1}^{(l)} - \widehat{\bar{W}}_{n+1}^{(l)} \right) \nonumber \\ 
& = \frac{1}{N} \sum_{n=1}^{N-1}  \frac{1}{L} \sum_{l=1}^L \mathbb E  \left[  \sum_{k=1}^{N} u_{n,k} \left(Y_k^{(l)} - \widehat{Y}_k^{(l)} \right) \sum_{p=1}^{N}  u_{n+1,p} \left( Y_p^{(l)} - \widehat{Y}_p^{(l)} \right) \right]   \nonumber  \\ 
& = \frac{1}{N} \sum_{n=1}^{N-1}  \frac{1}{L} \sum_{l=1}^L \sum_{k=1}^{N} u_{n,k} u_{n+1,k} \mathbb E  \left[  \left(Y_k^{(l)} - \widehat{Y}_k^{(l)} \right)^2 \right] \nonumber \\ 
& = \frac{1}{N} \sum_{n=1}^{N-1}  \sum_{k=1}^{N} u_{n,k} u_{n+1,k} \frac{1}{L} \sum_{l=1}^L \mathbb E  \left[  \left(B_l^{(k)} - \widehat{B}_l^{(k)} \right)^2 \right]. \label{eq:ce_proof_step_i}
\end{align}

\subsubsection*{Proof of Claim II}
We have
\begin{align}
& \frac{1}{N} \sum_{n=1}^{N-1} u_{k,n} u_{k,n+1} = \frac{A_k^2}{N} \sum_{n=1}^{N-1} \sin \left( \frac{2k-1}{2N+1} n \pi  \right) \sin \left( \frac{2k-1}{2N+1} (n+1) \pi  \right)  \nonumber \\
& = \frac{A_k^2}{2N} \sum_{n=1}^{N-1} \left(\cos\left( \frac{2k-1}{2N+1} \pi \right) - \cos \left( \frac{2k-1}{2N+1} \pi (2n+1) \right)  \right) \nonumber  \\
& = \frac{A_k^2}{2} \cos\left( \frac{2k-1}{2N+1} \pi \right) +o(1) \label{eq:ce_proof_sums}
\end{align}
where the last transition is since 
\[
\sum_{n=1}^{N-1} \cos \left( \frac{2k-1}{2N+1} \pi (2n+1) \right) 
\]
is bounded in $N$. From \eqref{eq:ce_proof_sums} we obtain:
\begin{align}
& \frac{1}{N} \sum_{n=1}^{N-1} \mathbb E \Delta_n \Delta_{n+1}  = \frac{1}{N} \sum_{n=1}^{N-1} \sum_{k=1}^N u_{k,n} u_{k,n} \min\left\{\theta, \lambda_k \right\} \nonumber \\
& =  \frac{1}{N} \sum_{k=1}^N (NA_k^2) \min\left\{\theta, \lambda_k \right\}  \frac{1}{N} \sum_{n=1}^{N-1}  u_{k,n} u_{k,n+1} \nonumber \\
& =  \frac{1}{2N} \sum_{k=1}^N \min\left\{\theta, \lambda_k \right\} (NA_k^2) \left(\cos\left( \frac{2k-1}{2N+1} \pi \right) +  O(1) \right).  \label{eq:ce_proof_sum2}
\end{align}
We now take the limit $N\rightarrow \infty$ as $k/N \rightarrow \phi$, so the spectrum of $\Sigma_{\bar{W}}$ converges to $S_{\bar{W}}(\phi)$. Moreover, since
\begin{align*}
A_k^{-2} & = \sum_{l=1}^N (u_{k,l})^2 = \sum_{l=1}^N  \sin^2 \left(\frac{2k-1}{2N+1} \pi l \right) \\
& = \sum_{l=1}^N \left( \frac{1}{2} - \frac{1}{2} \cos \left(2\frac{2k-1}{2N+1} \pi l \right) \right),
\end{align*}
we have $N A_k^2  \rightarrow 2$. Therefore, after multiplying by $\sigma^2/f_s$ to obtain the un-normalized distortion, \eqref{eq:ce_proof_sum2} converges to 
\begin{align*}
&  \frac{\sigma^2}{f_s} \int_0^1 \min \left\{ S_{\bar{W}}(\phi),\theta \right\} \cos(\pi \phi) d\phi  \\
& = D_{\bar{W}}(\bar{R}) -2 \frac{\sigma^2}{f_s} \int_0^1 \min \left\{ S_{\bar{W}}(\phi),\theta \right\}  \sin^2(\pi \phi /2) d\phi \\
& = D_{\bar{W}}(\bar{R}) - \frac{\sigma^2}{2f_s} \int_0^1 \frac{ \min \left\{ S_{\bar{W}}(\phi),\theta \right\} }{S_{\bar{W}}(\phi)}  d\phi \\
& = D_{\bar{W}}(\bar{R}) -  \frac{1}{2}G(\bar{R}).
\end{align*}

\subsubsection*{Proof of Claim III}
We have
\[
D_{\bar{W}}(\bar{R}) - \frac{1}{6}G(\bar{R}) = \int_0^1 \min \{ S_{\bar{W}}(\phi), \theta \} \left( 1- \frac{1}{6S_{\bar{W}}(\phi)} \right) d \phi,
\]
so 
\[
D_{\CE}(f_s,R) = \frac{1}{6f_s} + D_{\bar{W}}(\bar{R}) - \frac{1}{6}G(\bar{R}). 
\]

We now use Claims I-III to prove (i) and (ii) in Theorem~\ref{thm:ce_exact}. To show (ii), we fix $\bar{R} = R/f_s$ and 
consider the converse for the source coding theorem for encoding the $L$-dimensional vector source $B^{(k)}$, consisting of i.i.d.\ Gaussian random variables of variance $\lambda_k$, using $L\bar{R}_k$ bits. This converse implies that for any decoder $g$ of the form \eqref{eq:ce_proof_g}, 
\begin{align*}
 \frac{1}{L} \sum_{l=1}^L \mathbb E  \left[  \left(B_l^{(k)} - \widehat{B}_l^{(k)} \right)^2 \right] \geq \lambda_k 2^{-2 \bar{R}_k} = \min\{ \theta, \lambda_k\}. 
\end{align*}
Therefore, using I, 
\[
\frac{1}{NL} \sum_{n=1}^{LN} \mathbb E \Delta_n \Delta_{n+1} \geq \frac{1}{N} \sum_{n=1}^N \sum_{k=1}^N u_{n,k} u_{n+1,k} \lambda_k \min\{ \theta, \lambda_k\}. 
\]
In addition, we use the converse for the source coding theorem for the discrete-time Wiener process $\bar{W}_{[\cdot]}$ from \cite{berger1970information} to obtain
\begin{equation}
\mmse\left(\bar{W}^{LN} | \bar{f} \left( \bar{W}^{LN} \right) \right) \geq D_{\bar{W}}(\bar{R}). 
\end{equation}
It follows from II that for any $\epsilon>0$, there exists $T_0$ large enough such that, for any $T>T_0$, 
\[
\frac{1}{NL} \sum_{n=1}^{LN} \mathbb E \Delta_n \Delta_{n+1} + \epsilon/ 3 > D_{\bar{W}}(\bar{R}) - \frac{1}{2} G(\bar{R}),
\]
\[
\mmse\left( \bar{W}^{LN} | \bar{f} \left( \bar{W}^{LN} \right) \right)+ \epsilon / 3 > D_{\bar{W}}(\bar{R}), 
\]
and 
\[
\mmse(W^{LT} | \bar{W}^{LT}) + \epsilon / 3 \geq \frac{1}{6f_s}.
\]
Finally, from \eqref{eq:lemma_exact_lower} we obtain
\begin{align*}
& \mmse\left( W^{LT} | \bar{f}\left( \bar{W}^{LN} \right) \right) \geq \mmse(W^{LT} | \bar{W}^{LN}) \\
& \quad \quad + \frac{2}{3} \mmse\left( \bar{W}^{LN} | \bar{f} \left( \bar{W}^{LN} \right) \right) \\
& \quad > \frac{1}{6f_s} + \frac{2}{3} D_{\bar{W}}(\bar{R}) + \frac{1}{3} \left(D_{\bar{W}}(\bar{R}) - \frac{1}{2} G(\bar{R}) \right) - \epsilon.
\end{align*}
Since $L$ is arbitrary and using III, we conclude that for any $\epsilon>0$, there exists $T_0$ such that 
\[
\mmse \left(W^T |  \bar{f} \left( \bar{W}^N \right) \right) + \epsilon \geq D_{\CE}(f_s,R). 
\]

In order to prove (i), fix $\rho,\epsilon, \delta>0$, and consider a decoder $g$ that, upon receiving $(i_1,\ldots,i_{n_{\max}})$, first computes the inverse transform $\Um^T \widehat{b}(i_n)$ for each index $i_n$ and concatenates the resulting vectors to obtain $\widehat{\bar{W}}^{LN_T}$. In order to estimate $W^{TL}$, the decoder uses an interpolation similar to \eqref{eq:W_tilde_def}:
\begin{equation}
\label{eq:linear_interp_ce}
\widehat{W}_t \triangleq \frac{t^+-t}{T_s} \widehat{\bar{W}}_{t^-} + \frac{t-t^-}{T_s} \widehat{\bar{W}}_{t^+}, \quad t\in [0,TL]. 
\end{equation}


In order to analyze the distortion resulting from using this decoder, consider first the $L$ dimensional vector $B^{(k)}$ using $\bar{R}_k+\rho$ codewords drawn i.i.d.\ from $\mathcal N(0, [\lambda_k - \theta]^+)$, where 
\[
\bar{R}_k = \begin{cases}  \frac{1}{2} \log [ \lambda_k / \theta], & k\leq k_{\max} \\
0 & k > k_{\max}. 
\end{cases}
\]
For any $T>0$ there exists $L_0$ that is independent of $T$, such that for any $k=1,\ldots, k_{\max}$ and $L>L_0$,
\begin{equation}
\frac{1}{L} \sum_{l=1}^L \mathbb E \left(B^{(k)}_l -\widehat{B}^{(k)}_l \right)^2 - \epsilon/3 \leq \lambda_k 2^{-2R_i} = \theta. 
\label{eq:ce_proof_Bi}
\end{equation}
Substituting \eqref{eq:ce_proof_Bi} in \eqref{eq:ce_proof_step_i}, we conclude that for $L \geq L_0$, 
\begin{align*}
& \frac{1}{NL} \sum_{n=1}^{LN} \mathbb E \Delta_n \Delta_{n+1} - \epsilon/3 \leq \frac{1}{N} \sum_{n=1}^N \sum_{k=1}^N u_{n,k} u_{n+1,k} \min \{ \lambda_k, \theta\}.
\end{align*}
Next, let $T_0$ be such that for all $T>T_0$,
\[
\left| \mmse \left( W^{L_0T} | \bar{W}^{L_0N} \right) -  \mmse(f_s) \right| < \epsilon / 3.
\]
Using the achievability side of the source coding theorem with respect to $\bar{W}_{[\cdot]}$ from \cite{berger1970information}, we may choose $T>T_0$ and $L = \sqrt{T} > L_0$ such that
\[
 \mmse\left(\bar{W}^{LN} | \bar{f} \left( \bar{W}^{LN} \right) \right) < D_{\bar{W}}(\bar{R}) + \epsilon/3,
\]
and therefore
\[
\mmse\left( W^{LT} | \bar{f} \left( \bar{W}^{LN} \right) \right) \leq D_{\CE}(R,f_s) + \epsilon. 
\]


\section*{Acknowledgments}
This work is supported in parts by NSF Center for Science of Information (CSoI) under grant CCF-0939370 and NSF-BSF under grant 1609695.

\bibliographystyle{IEEEtran}
\bibliography{/Users/kipnisal/LaTex/bibtex/IEEEfull,/Users/kipnisal/LaTex/bibtex/sampling}

\end{document}